\documentclass{sig-alternate-05-2015}
\usepackage[T1]{fontenc}
\usepackage{amsmath}
\usepackage{array}
\usepackage{graphicx}
\usepackage{color}
\usepackage[small,bf]{caption}
\usepackage[labelformat=simple,skip=1pt,margin=0pt,font=scriptsize]{subcaption}

\usepackage{times}
\usepackage{indentfirst}
\usepackage{url}
\definecolor{darkgreen}{RGB}{47,109,79}
\definecolor{darkblue}{RGB}{57,79,99}
\usepackage[bookmarks=false, colorlinks=true, plainpages = false,  linkcolor = darkgreen,   citecolor = darkgreen, urlcolor = darkblue, filecolor = blue]{hyperref}

\newcommand{\descr}[1]{\smallskip\noindent \textbf{#1}}

\begin{document}

\makeatletter\def\@copyrightspace{\relax}\makeatother
\setcopyright{acmcopyright}

\newfont{\mycrnotice}{ptmr8t at 7pt}
\newfont{\myconfname}{ptmri8t at 7pt}
\let\crnotice\mycrnotice%
\let\confname\myconfname%

\CopyrightYear{2016}
\makeatletter\def\@copyrightspace{\relax}\makeatother
\setcopyright{acmcopyright}
\conferenceinfo{SIGSPATIAL'16,}{October 31-November 03, 2016, Burlingame, CA, USA}
\isbn{978-1-4503-4589-7/16/10}\acmPrice{\$15.00}
\doi{http://dx.doi.org/10.1145/2996913.2996971}

\clubpenalty=10000 
\widowpenalty = 10000

\title{Privacy-Friendly Mobility Analytics using\\ Aggregate Location Data} 

\numberofauthors{3} 
\author{
\alignauthor{
Apostolos Pyrgelis\\
\affaddr{University College London}\\
\affaddr{apostolos.pyrgelis.14@ucl.ac.uk}}
\and
\alignauthor{
Emiliano De Cristofaro\\
\affaddr{University College London}\\
\affaddr{e.decristofaro@ucl.ac.uk}}
\and
\alignauthor{
Gordon J. Ross\\
\affaddr{University College London}\\
\affaddr{gordon.ross@ucl.ac.uk}}}

\maketitle

\begin{abstract}
Location data can be extremely useful to study commuting patterns and disruptions, as well as to predict real-time traffic volumes. At the same time, however, the fine-grained collection of user locations raises serious privacy concerns, as this can reveal sensitive information about the users, such as, life style, political and religious inclinations, or even  identities. In this paper, we study the feasibility of crowd-sourced  mobility analytics over {\em aggregate} location information: users periodically report their location, using a privacy-preserving aggregation protocol, so that the server can only recover aggregates -- i.e., how many, but not which, users are in a region at a given time. We experiment with real-world mobility datasets obtained from the Transport For London authority and the San Francisco Cabs network, and present a novel methodology based on time series modeling that is geared to forecast traffic volumes in regions of interest and to detect mobility anomalies in them. In the presence of anomalies, we also make enhanced traffic volume predictions by feeding our model with additional information from correlated regions. Finally, we present and evaluate a mobile app prototype, called Mobility Data Donors (MDD), in terms of computation, communication, and energy overhead, demonstrating the real-world deployability of our techniques.
\end{abstract}%

%
%\category{H.2.8}{Database Management}{Database Applications}[Spatial databases and GIS]
%\category{K.4.1}{Computers and Society}{Public Policy Issues}[Privacy]

%
%\keywords{\noindent Mobility Analytics, Location Privacy, Data Aggregation}

%
\section{Introduction}
\label{sec:intro}
The availability of information about people's locations and movements holds the promise to make urban planning more effective and efficient, and ultimately improve citizens' quality of life. Prompted by the increased presence of always-on, always-connected devices, the ubiquitous collection of location information enables a number of interesting applications. 
New research frontiers, e.g., in the field of \textit{anticipatory mobile computing}, make it increasingly possible to use mobile sensing along with machine learning for intelligent reasoning~\cite{pejovic2015anticipatory}.  
For instance, contextual location information collected from mobile users can be used to predict future mobility events~\cite{lam2005short,stathopoulos2003multivariate}, detect mobility anomalies~\cite{pan2013crowd} and enable real-time traffic or event statistics~\cite{waze}. %

At the same time, however, large-scale collection of individual users' fine-grained locations raises serious privacy concerns, 
as this can reveal sensitive information about the users, such as, life style, political and religious inclinations, or even  identities~\cite{taxis-rainbows,krumm2007inference}. %
Although often advocated, anonymization of location traces is moot as these reveal home/work locations, which in turn can be used to re-identify users~\cite{golle2009anonymity}. In fact, just a few locations are enough to re-identify users~\cite{zang2011anonymization}. 
Therefore, in this paper, we set to investigate whether or not mobility analytics can be effectively and efficiently performed over aggregate data.
We turn to cryptographic protocols for privacy-friendly data aggregation and use them to privately gather location statistics~\cite{brown2013haze,kopp2012privacy,melis,popa2011privacy}. %
Overall, we aim to demonstrate: (1) the usefulness of mobility analytics over aggregate locations, and (2) the real-world deployability of a scalable system for privacy-friendly location data collection. 

\descr{Roadmap.} We present a crowd-sourced system for privacy-friendly mobility analytics whereby users periodically report locations, but do so using a privacy-preserving aggregation protocol, so that only aggregates can be recovered (i.e., how many but not which users were in a region at a given time). We experiment with real-world mobility datasets obtained from the Transport For London (TFL) authority as well as the San Francisco Cabs (SFC) network, and present a methodology based on time series modeling geared to forecast traffic volumes in regions of interest (ROIs) and to detect mobility anomalies in them. In the presence of anomalies, we also make enhanced traffic volume predictions (achieving up to $50\%$ improvement) by training our model with additional information from correlated regions. Such tasks are particularly useful in modern cities for journey planning~\cite{waze,lam2005short} and congestion prevention~\cite{silva2015predicting}.
Finally, we show how to build a privacy-respecting system for data collection. To this end, we present a mobile app prototype, called Mobility Data Donors (MDD), and present an empirical evaluation of its computation, communication, and energy complexities, which attest to the practicality of our vision.

\descr{Paper Organization.} 
Next section introduces a few concepts and tools used in our work, then, Section~\ref{sec:analytics} presents our datasets and our methodology for predictive mobility analytics. In Section~\ref{sec:system}, we discuss the details of our proposed framework and analyze its real-world deployment. Finally, after reviewing related work in Section~\ref{sec:related}, the paper concludes in Section~\ref{sec:concl}.

\section{Preliminaries}
\label{sec:prelim}

%
%
%%\vspace*{-0.1cm}
\subsection{Auto Regressive Moving Average}\label{sec:arma}
As we aim to perform analytics on aggregate locations -- specifically, predicting traffic volumes as well as detecting mobility anomalies in a Region Of Interest (ROI), such as underground stations (cf.~Section~\ref{sec:predictions}--\ref{sec:anoms}) -- we model ROIs' time series using Auto-Regressive Moving Average (ARMA). We build on the work by Box et al.~\cite{box2015time}, who 
present an iterative method for choosing and estimating ARMA models. 

Given a time series $Y_{t}$, an ARMA model is a tool for understanding and predicting future values in $Y_{t}$.
The model is usually denoted as $ARMA(p, q)$, where $AR(p)$ denotes the autoregressive model of order p
and $MA(q)$ refers to the moving average model of order q. Specifically, an $ARMA(p,q)$ model is defined as:

%%\vspace*{-0.45cm}
\begin{equation}
\label{eq:ARMA}
Y_t = c +  \sum_{i=1}^{p} \phi_i \cdot Y_{t - i} + \epsilon_t + \sum_{i=1}^{q} \theta_i \cdot \epsilon_{t - i}
%%\vspace*{-0.15cm}
\end{equation}
where $c$ is a constant, $\phi_1, \ldots, \phi_p$ and $\theta_1, \ldots, \theta_q$ are model parameters, and $\epsilon_{t}, \epsilon_{t-1}, \ldots$ are white noise error terms.

\subsection{Vector Auto-Regression}
We also investigate how to improve traffic volume predictions in the presence of anomalies (cf.~Section~\ref{sec:improv}), thus, we also attempt to discover correlated ROIs %
and use their aggregate time series, along with a Vector Auto-Regression (VAR) model, to make enhanced predictions. VARs are statistical models used in econometrics to capture linear interdependencies among multiple time series, and consist a generalization of uni-variate autoregressive models (AR models) that allow more than one evolving variable. All variables in a VAR model are treated symmetrically and each of them has an equation explaining its evolution based on its own lags as well as those of the other model variables. VAR modeling requires the prior knowledge of a list of variables which can be hypothesized to affect each other inter-temporally.

A VAR model describes the evolution of a set of $k$ variables (endogenous variables) over a sample period $t = 1, \ldots, T$ as a linear function of their past values. The variables are collected in a vector $y_{t}$ of size $(k, 1)$, whose $i_{th}$ element $y_{it}$ is the observation of the variable $i$ at time $t$. A $p$-th order VAR model, denoted as $VAR(p)$ is given by the equation:%%\vspace*{-0.1cm}
\begin{equation}
y_t = c  + A_{1} \cdot y_{t-1} + A_{2} \cdot y_{t-2} + \ldots A_{p} \cdot y_{t-p} + e_{t}%%\vspace*{-0.1cm}
\end{equation}
where $c$ is a vector of constants with size $(k, 1)$, $A_{i}$ is a time-invariant matrix of size $(k, k)$ and $e_t$ is a vector of error terms with size $(k, 1)$ where: (a) $E(e_t) = 0 $, every error term has mean zero, (b) $E(e_{t} e^{'}_{t}) = \Omega$, the co-variance matrix of error terms is $\Omega$ and (c) $E(e_{t}e^{'}_{t-k}) = 0$, for any non-zero $k$ there is no serial correlation in individual error terms.

\subsection{Spearman Correlation}
To discover correlated ROIs, %
we will use Spearman's correlation coefficient, which is a non-parametric measure of the statistical dependence between the ranking of two variables~\cite{corder2014nonparametric}. It provides an estimate of how well the relationship between two variables can be described with a monotonic function and, unlike Pearson, it does not assume that both variables are normally distributed. 
Given two variables $W, Z$, the Spearman correlation coefficient is defined as: %%\vspace*{-0.2cm}
\begin{equation}
r_s = 1 - \frac{6 \cdot \sum d^{2}_i }{n \cdot (n^{2} - 1)}
\end{equation} 
where $d_i = rg(W_i) - rg(Z_i)$ is the difference between the two ranks of each observation and $n$ is the number of observations. Similar to other correlation measures, Spearman's obtains values between $-1$ and $+1$, with $0$ implying no correlation, and $-1$ or $+1$ implying an exact monotonic relationship. Intuitively, positive correlations imply that as $W$ increases, so does $Z$, while negative correlations mean that as $W$ increases, $Z$ decreases.

%%\vspace*{-0.1cm}
\subsection{Privacy-Preserving Data Aggregation}\label{sec:aggr}
We also use cryptographic protocols for privacy-preserving data aggregation, allowing an untrusted aggregator to gather statistics (e.g., sum or mean) from  users in such a way
that data of single users is not revealed in the clear, but only the aggregate information can be recovered. %
These protocols are often used for smart metering~\cite{kursawe2011privacy}, participatory sensing~\cite{popa2011privacy}, or recommender systems~\cite{melis}.

Typically, private aggregation relies on a cryptosystem that is additively homomorphic: users send encrypted data to the aggregator, which does not hold the corresponding decryption key and cannot 
access {\em single} users' contributions, however, it can decrypt the sum of {\em all} users' reports.
Specifically, we choose a protocol recently proposed by Melis et al.~\cite{melis}, as it guarantees: scalability, independence from trusted third parties and/or key distribution centers, and fault tolerance. Scalability is achieved by combining the private aggregation protocol of Kursawe et al.~\cite{kursawe2011privacy} (secure under the Computational Diffie Hellman assumption %
in the presence of honest-but-curious adversaries) with data structures supporting succinct data representation, i.e., Count-Min Sketches~\cite{cormode2005improved}. These introduce a small, upper-bounded error in the aggregation, but reduce the computational/communication complexities of the cryptographic operations from linear to logarithmic in the size of the input. It also features a completely distributed key generation/distribution which, unlike other protocols, e.g.~\cite{jawurek2012fault,popa2011privacy}, does not require any other authorities. Finally, its  fault tolerance
protocol addresses one of the main limitations of~\cite{kursawe2011privacy}, i.e., if one or more users fails to report their (encrypted) data, the aggregator cannot correctly decrypt the aggregate (since it relies on encryption keys summing up to zero).

Melis et al.~\cite{melis}'s protocol consists of four phases. {\bf\em (1) Setup:} Assuming a cyclic group $\mathbb{G}$ of order $q$ for which the Computational Diffie-Hellman problem is hard, and $g$ a generator of this group, each user $U_i \in \mathbb{U} = \{1, \ldots, N \} $ generates a private key $x_{i} \in_{r} \mathbb{G}$ (i.e., sampled at random from $G$) and a public key $y_{i} = g^{x_{i}} \, mod \, q$. The public keys are published with the aggregator. 
{\bf\em (2) Encryption:} Each user $U_i$ holds an input vector of data points $S = \{ S_c \in \mathbb{N}, c = \{1, \ldots, T \} \}$. %
To participate in the privacy-preserving aggregation each user needs to generate blinding factors based on the public keys of the other users in such a way that they all sum up to zero. At round $s$, for $l = 1, \ldots, T$, user $U_i$ computes $k_{il} = \sum_{\substack{j=1, j \neq i}}^{N} H(y_{j}^{x_{i}} \| l \| s ) \cdot (-1)^{i > j} \, mod \, q$, where $H$ %
is a cryptographic hash function and $\|$ denotes the concatenation operator. Then, for each entry $\{S_{il}\}_{l=1}^{T}$,  $U_i$ encrypts $S_{il}$ as $b_{il} = S_{il} + k_{il} \, mod \, 2^{32}$ and sends the resulting ciphertext to the aggregator. 
{\bf\em (3) Aggregation:} The aggregator collects the ciphertexts from each user $U_i$ and (obliviously) aggregates them. More precisely, for $l = 1, \ldots, T$ it computes $C_l = \sum_{i = 1}^{N} b_{il}  = \sum_{i =1}^{N} k_{il} + \sum_{i=1}^{N} S_{il}  = \sum_{i=1}^{N} S_{il} \, mod \, 2^{32}$, where $C_l$ denotes the $l-th$ item of the input vector $S$. 
{\bf\em (4) Fault Recovery:} If, during the aggregation phase, only a subset of users $\mathbb{U}_{on}$ successfully submit data, the aggregator sends $\mathbb{U}_{on}$ to each $U_i \in \mathbb{U}_{on}$ and $U_i$ computes, for each $l = 1, \ldots, T$, $k^{'}_{il} = \sum_{\substack{j=1, j \neq i, j \notin \mathbb{U}_{on}}}^{N} H(y_{j}^{x_{i}} \| l \| s ) \cdot (-1)^{i > j} \, mod \, q$. Then each user $U_i$ sends these values back to to the aggregator who can now obtain the aggregate counts by computing $C^{'}_{l} = ( \sum_{i \in \mathbb{U}_{on}}^{} b_{il} - \sum_{i \in \mathbb{U}_{on}}^{} k^{'}_{il} ) \, mod \, 2^{32}$.

\descr{Groups.} Another feature of~\cite{melis} is the ability to dynamically allocate users in groups, and perform within-group aggregation and then combining statistics from multiple groups, which is crucial to cope with dynamic/mobile settings.
It also allows to bound the complexity of the encryption phase, which depends on the number of users in the group.

\descr{Input Compression.} As mentioned above,~\cite{melis} uses Count-Min Sketches to guarantee scalability when the input vector ($S$) is large. Specifically, the encryption phase is modified as follows: each user $U_{i}$ initializes a Count-Min Sketch vector $X_{i} \in \mathbb{N}^{d \times w}$ with zero entries, then encodes his original input vector $S$ using the update procedure of Count-Min Sketches~\cite{cormode2005improved} while employing the following pairwise hash function: $h(x) = ( (a \cdot x + b) \, mod \, p ) \, mod \, w$ for $a \neq 0, b$ random integers modulo a random prime $p$. Then each user encrypts $X_{i}$ as in the previously described encryption phase.

If $\vert S \vert$ denotes the size of the input vector $S$, its compact representation with a Count-Min Sketch has size $O(\log(\vert S \vert))$. More precisely, given the sketch parameters $(\epsilon, \delta)$, the Count-Min Sketch is a vector of size $L= d \times w$ where $d=\lceil \ln{(\vert S \vert / \delta)} \rceil$ and $w = \lceil e / \epsilon \rceil$. For instance, if $\epsilon = \delta = 0.01$, a vector $S$ of size $\vert S \vert = 10 ^ 4$ can be encoded as a sketch of size $L = 3,808$, while a vector $S$ of size $\vert S \vert = 10^6$ can be represented as a sketch of size $L = 5,168$. Obviously, the Count-Min succinct structure introduces an accuracy error and its parameters $(\epsilon, \delta)$ give an upper bounded error for the estimated counters $\hat{c_i}$, amounting to 
$\hat{c_i} \leq c_i + \epsilon \cdot \sum_{j} |c_j|$ with probability $1-\delta$ (with $c_i$ being the true element of the vector).

\section{Mobility Analytics using\\ Aggregate Locations}
\label{sec:analytics}

We now present and evaluate our ``mobility analytics'' algorithms, specifically, predicting traffic volumes at ROIs, discovering and predicting anomalies -- all using aggregate location reports. We rely on two real-world datasets obtained, respectively, from Transport for London (TFL) and the San Francisco Cab (SFC) network.

\subsection{Datasets}
\label{sec:datasets}

%

%%\vspace*{-0.15cm}
\subsubsection{Transport For London (TFL)}
London's transportation system consists of various connected subsystems: London Underground Ltd (LUL), London Transport Buses (LTB), Docklands Light Rail (DLR), Overground (LRC), Tramlink (TRAM), and National Rail (NR), %
operating in the city under the umbrella of Transport for London (TFL). The most common payment method for TFL fares is the Oyster Card, a pre-paid, RFID-enabled card.
We have obtained from TFL data corresponding to all March 2010 trips from all (anonymized) oyster cards, which we pre-process in the following way. First, we discard trips from TRAM due to scarce density and LTB for consistency as travelers only tap-in but do not tap-out for bus trips paid by Oyster.
Then,  to observe weekly patterns, we focus on the four weeks from Monday March 1 to Sunday 28, 2010. The final dataset consists of approximately 60 million oyster-card trips, performed by almost 4 million unique users, over 582 stations.
Each entry in the data describes a unique trip and consists of the following fields: \textit{oyster id, start time, start station id, end time,} and \textit{end station id}. %
Note that the time resolution of the timestamps is 1 minute. 

Next, we aggregate single-trip records by grouping trips start and end times in time epochs of 1 hour, aiming to achieve regularity in transit patterns (similar to~\cite{zhong2016variability}), and count the number of passengers that entered (``tap-in'') or exited (``tap-out'') each station during a slot. For each station $i  \in \{1, \ldots, n\}$ (with $n$ being the total number of stations), we create a time series $Y_{it}$ indicating how many passengers transited through it in a time epoch $t \in \{1, \ldots, m\}$ ($m$ denotes the total number of epochs, i.e., 672): %
$Y_{it} = Y^{in}_{it} + Y^{out}_{it}$, %
where $Y^{in}_{it}$ indicates the number of tap-in events and $Y^{out}_{it}$ the number of tap-out events, at station $i$ during epoch $t$. 

\begin{figure}[t]
\centering
   \begin{subfigure}{0.355\textwidth}
   \includegraphics[width=1\linewidth]{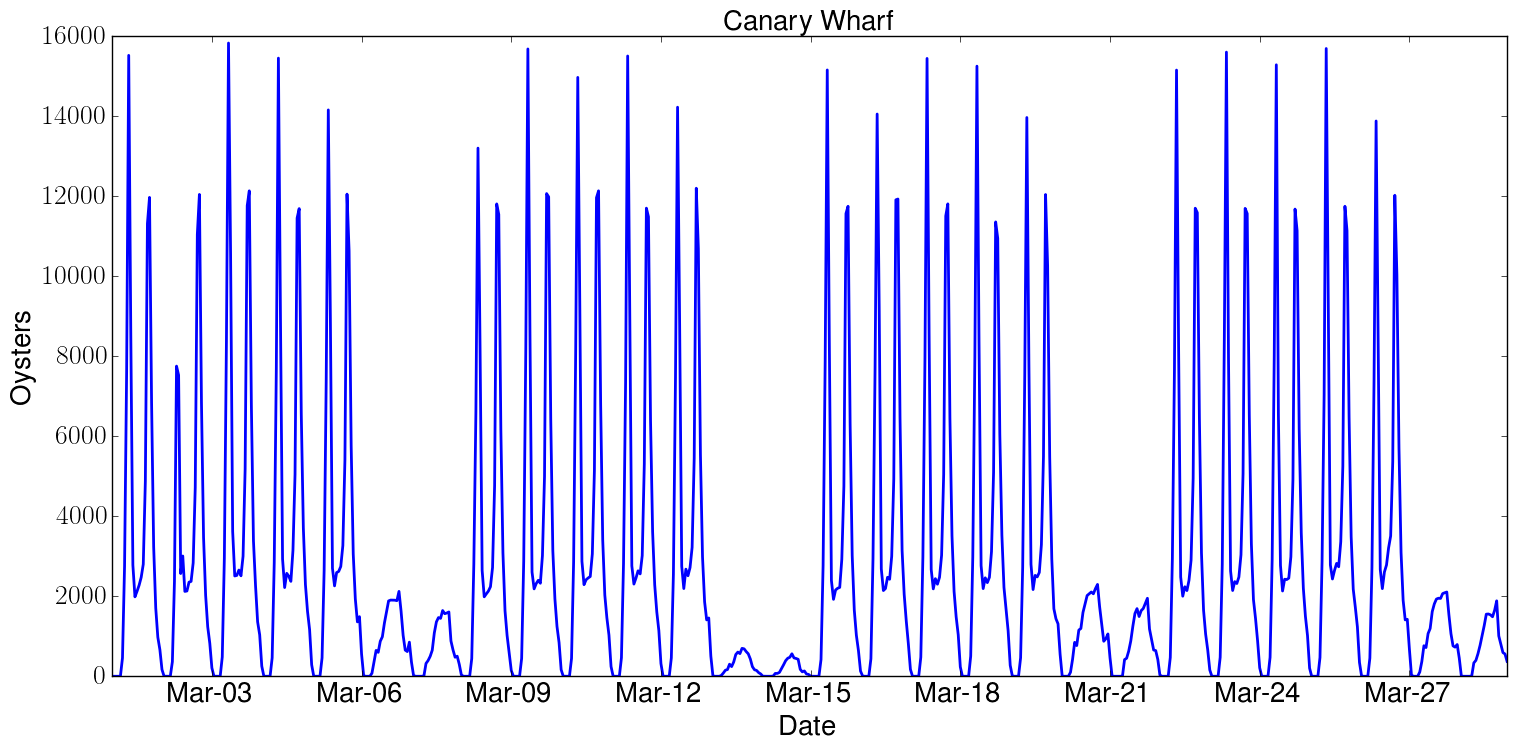}
   \caption{}
   \label{fig:canary-wharf-ts} 
\end{subfigure}
\begin{subfigure}{0.355\textwidth}
   \includegraphics[width=1\linewidth]{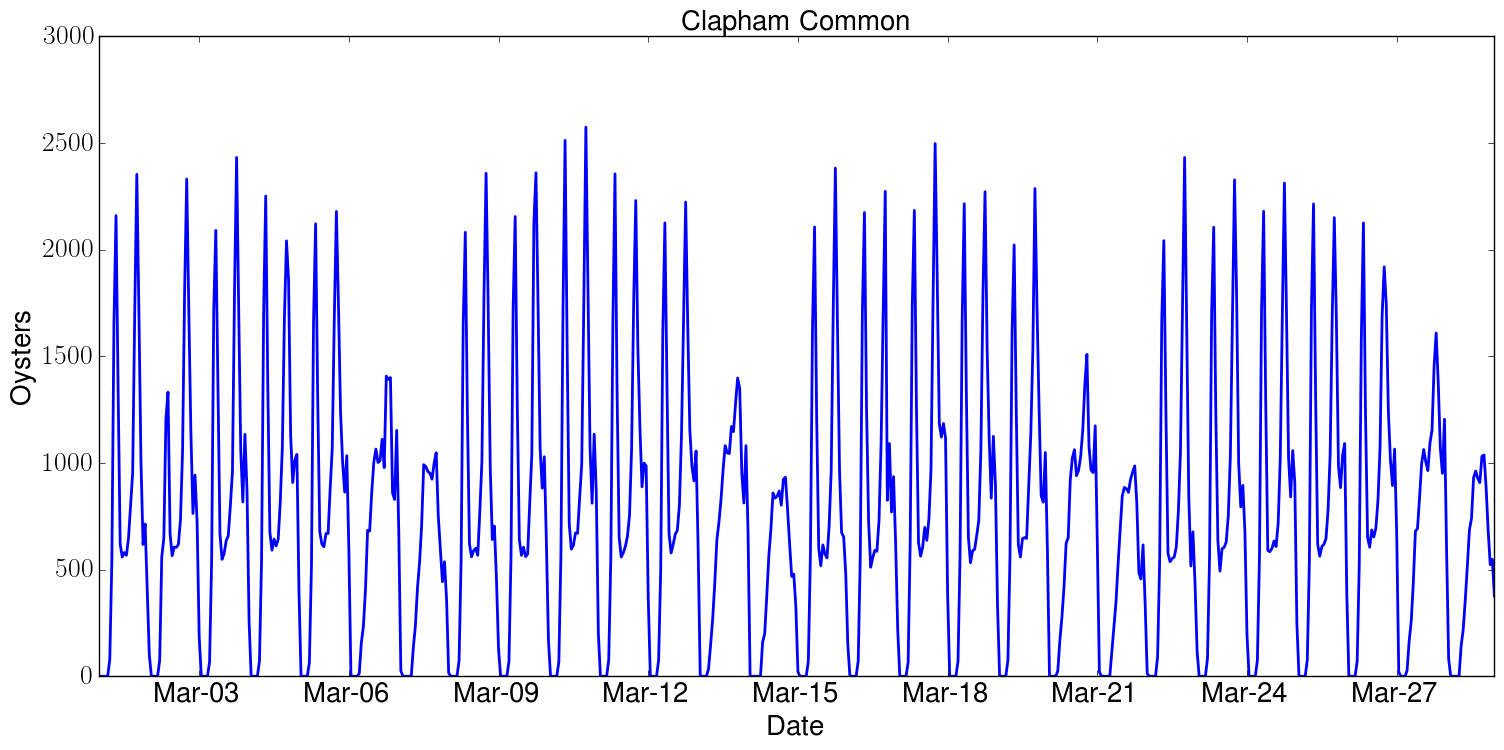}
   \caption{}
   \label{fig:clapham-ts}
\end{subfigure}
%%\vspace*{-0.3cm}
\caption{Hourly traffic volume at two TFL stations (March 1--28, 2010).}
%%\vspace*{-0.3cm}
\label{fig:various}
\end{figure}

In Figure~\ref{fig:various}, we plot the \textit{hourly} aggregate time series of two stations -- Canary Wharf (one of the busiest stations of London) and Clapham Common (a moderately busy station) -- showing different patterns during weekdays and weekends, as well as peak commuting hours. In general, we note some weekly/daily seasonality in the stations' time series as well as stationarity (i.e., no particular trend). We verify the latter by performing the Augmented Dickey-Fuller test~\cite{dickey1979distribution} which indicates that $93\%$ of tube stations have stationary time series with $95\%$ confidence. 

%%\vspace*{-0.1cm}
\subsubsection{San Francisco Cab (SFC)}
We also use the San Francisco Cab (SFC) dataset~\cite{epfl-mobility-20090224}, which contains mobility traces recorded by taxis in San Francisco, between May 17 to June 10, 2008. The dataset contains approximately 11 million GPS coordinates, generated by 536 taxis. To observe weekly patterns in our data we sample the dataset to cover exactly 3 weeks of data: Monday May 19 to Sunday June 8, 2008. Entries in the dataset %
include the following fields: \textit{cab identifier, latitude, longitude} and a \textit{time stamp} in UNIX epoch format.

We follow a similar approach as with the TFL dataset to aggregate the traces, however, since locations are GPS coordinates rather than points of interest, we divide the city of San Francisco into a grid $S$ consisting of $100\times 100$ regions, each covering an area of $ 0.19 \times 0.14$ square miles. We group the GPS traces in one-hour  epochs and, for each region $i \in \{1, \ldots, n\}$ (with $n$ being the total number of regions, i.e., 10,000), we count the number of taxis that have reported a presence in that block during epoch $t \in \{1, \ldots, m \}$ ($m$ being the number of time epochs, i.e., 504), and create a time series $Y_{it}$ as: %
$Y_{it} = \sum_{j=1}^{k} p_{jt}$, %
where $k$ is the total number of taxis (i.e., $536$) and $p_{jt} \in \{0, 1\}$ indicates whether taxi $j \in \{1, \ldots, k \}$ reported its location at region $i$ during epoch $t$.

\begin{figure}[t]
\centering
   \includegraphics[width=0.65\linewidth]{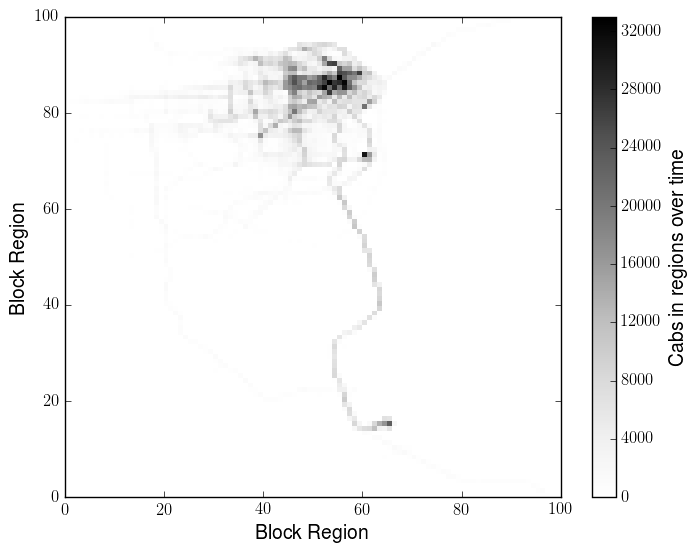}
   %%\vspace*{-0.3cm}
   \caption{Number of cabs on $100\times 100$ SF grid (May 19 -- June 8, 2008).}
\label{fig:SF-grid}
\end{figure}

\begin{figure}[t]
\centering
\begin{subfigure}[b]{0.355\textwidth}
   \includegraphics[width=1\linewidth]{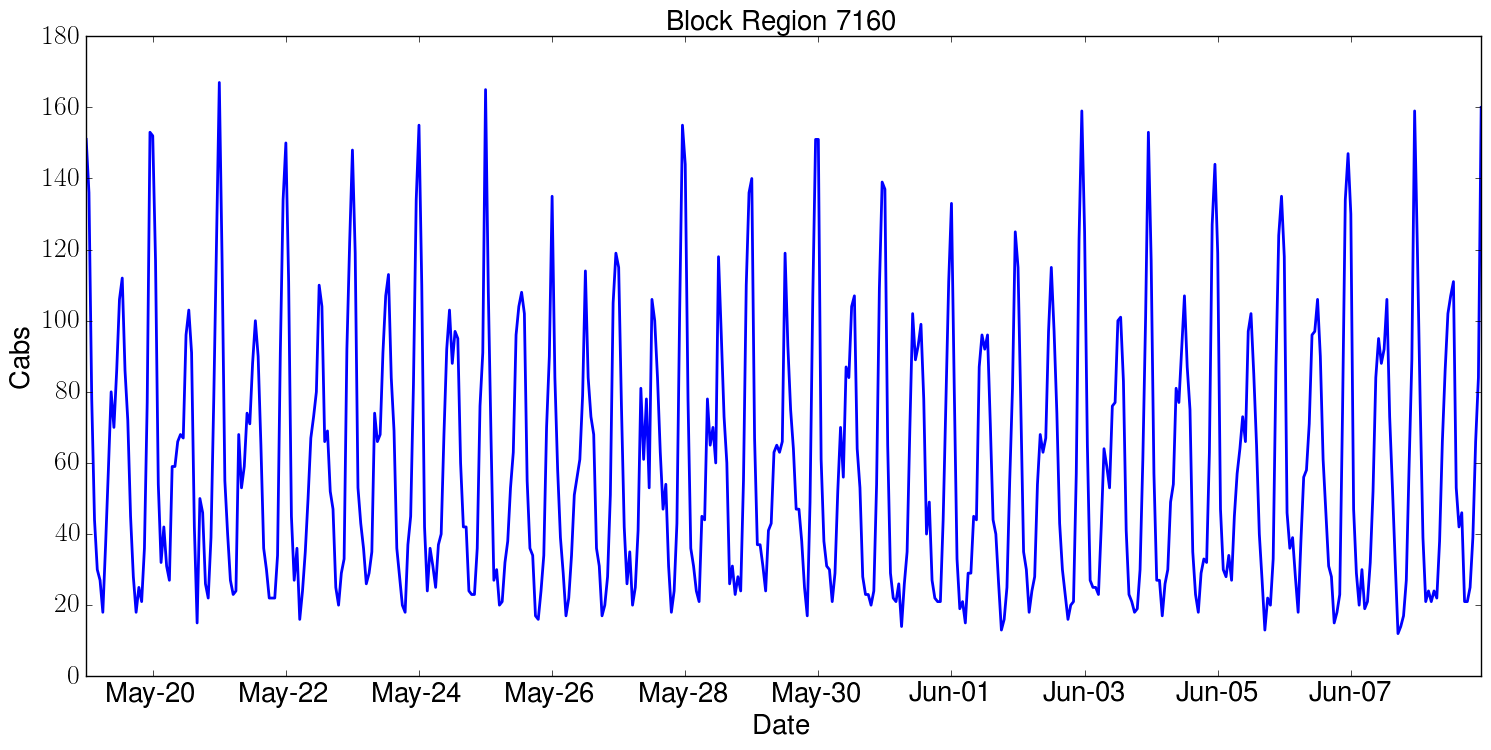}
   \caption{}
   \label{fig:region-10-ts} 
\end{subfigure}
\begin{subfigure}[b]{0.355\textwidth}
   \includegraphics[width=1\linewidth]{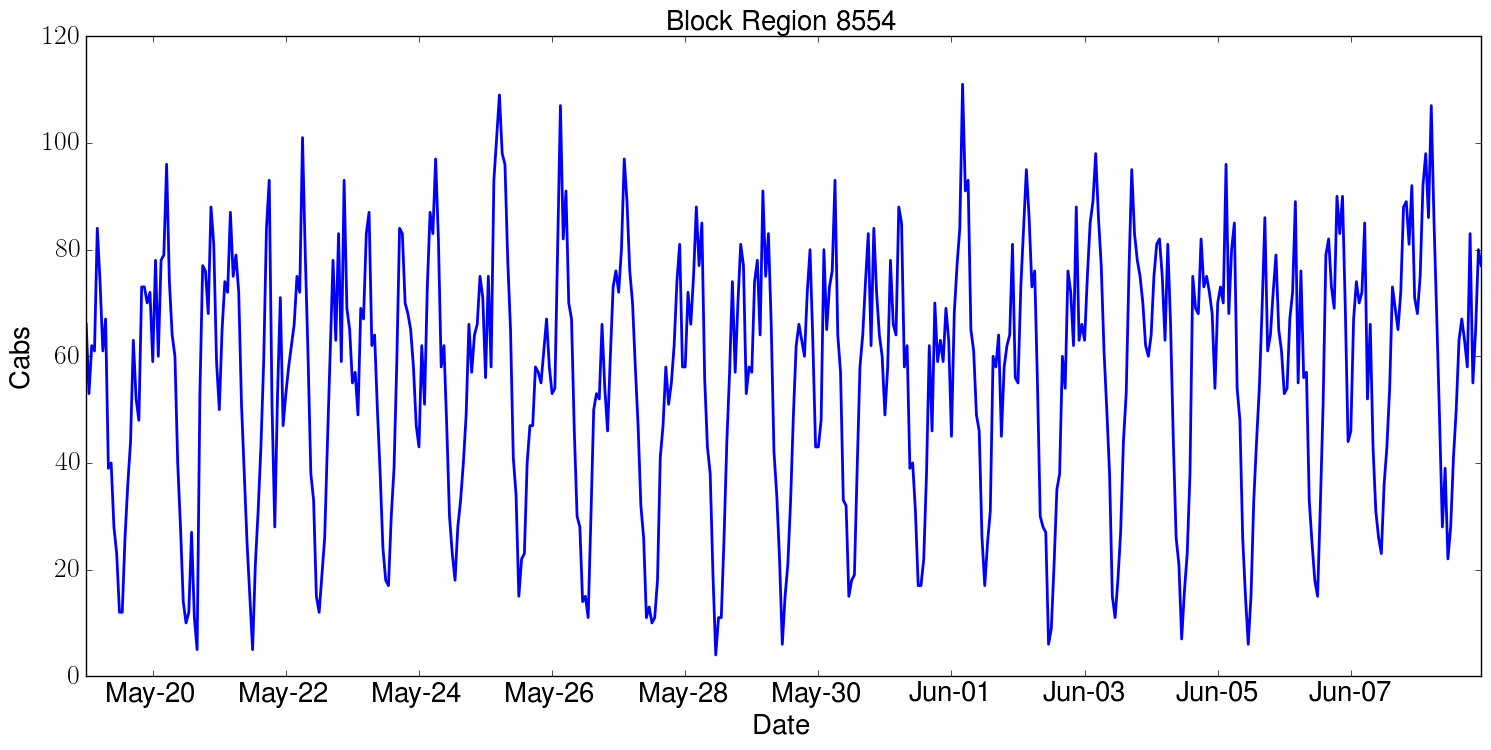}
   \caption{}
   \label{fig:region-11-ts} 
\end{subfigure}
%%\vspace*{-0.3cm}
\caption{Hourly traffic volume in regions 7160 and 8554 of SFC dataset.}
\label{fig:sfc-regions-ts}
%%\vspace*{-0.3cm}
\end{figure}

We aggregate the traffic from our 3-week dataset for each region and, in Figure~\ref{fig:SF-grid}, plot the resulting heatmap. Unsurprisingly, the downtown area exhibits the highest traffic volume, with the route to/from SFO airport also clearly visible. %
In Figure~\ref{fig:sfc-regions-ts}, we plot the aggregate time series of two regions, one of the busiest (id = 7160) and a moderately busy one (id = 8554). Once again, weekly and daily patterns can be observed, along with stationarity (96 out of the 100 busiest regions have stationary time series with $99\%$ confidence as indicated by the Augmented Dickey-Fuller test). %

%
%
%%\vspace*{-0.1cm}
\subsubsection{Removing Seasonality}\label{sec:seasonal}
\begin{figure}[t]
\centering
   \begin{subfigure}[b]{0.355\textwidth}
   \includegraphics[width=1\linewidth]{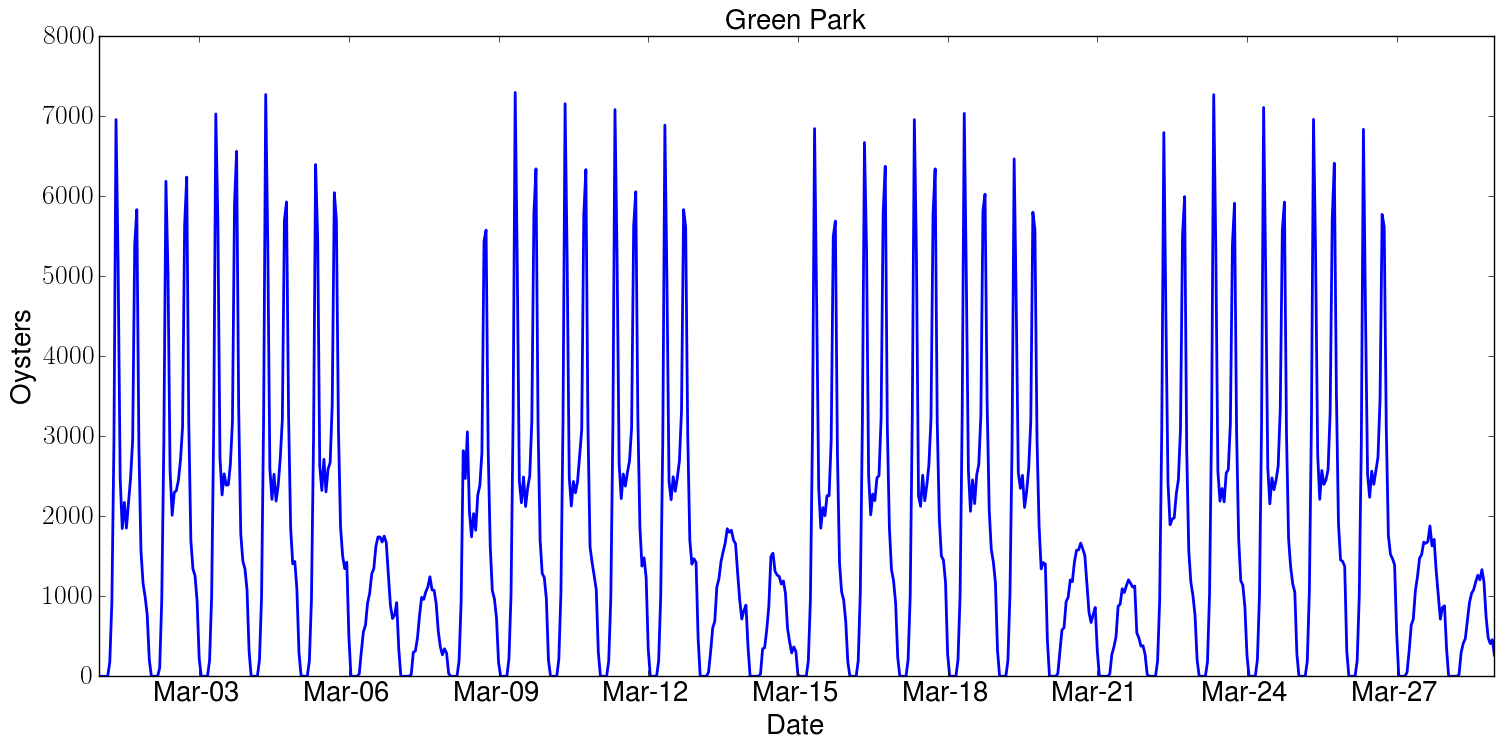}
   \caption{}
   \label{fig:green-park-ts} 
\end{subfigure}
\begin{subfigure}[b]{0.355\textwidth}
   \includegraphics[width=1\linewidth]{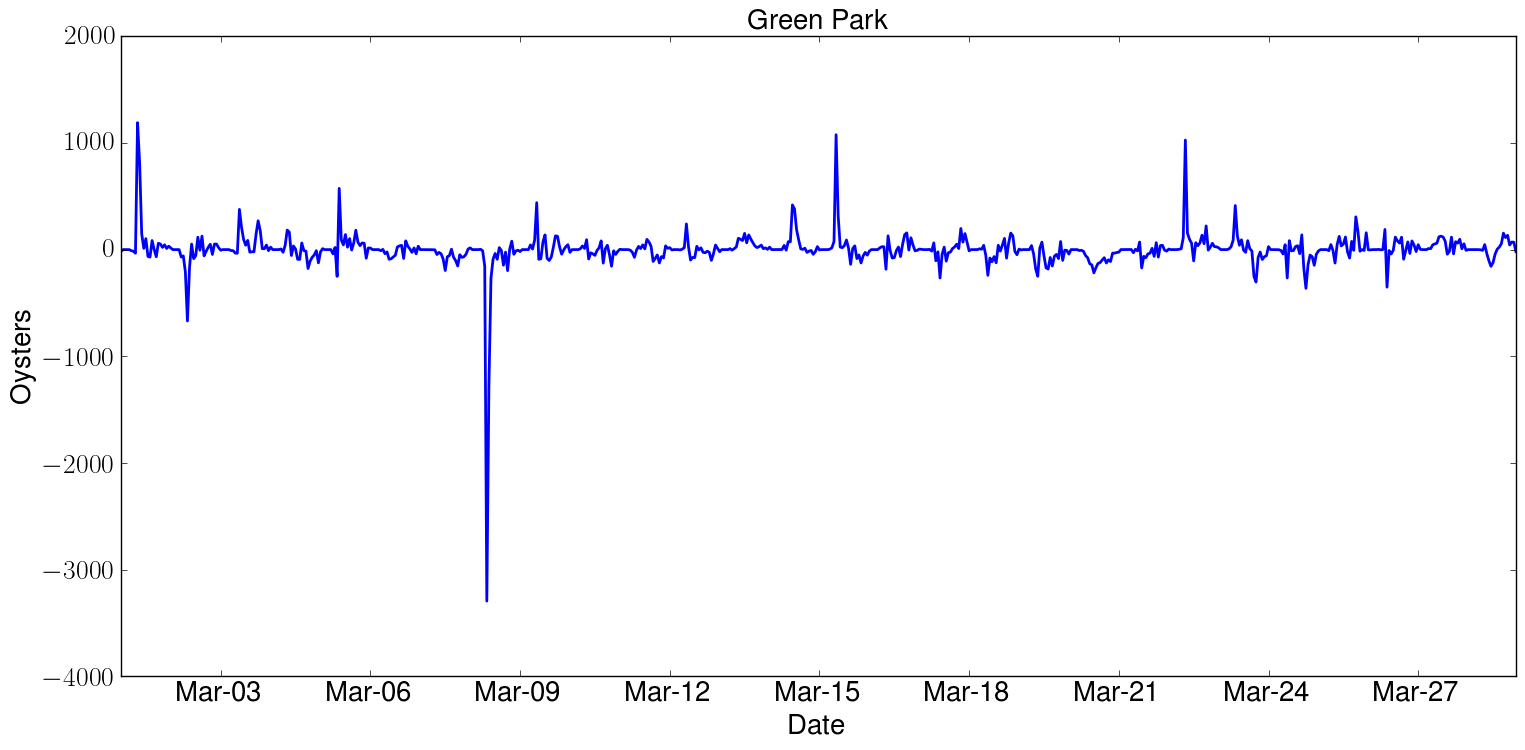}
   \caption{}
   \label{fig:green-park-des-ts}
\end{subfigure}
%%\vspace*{-0.3cm}
\caption{Green Park station's time series without (a) and with (b) de-seasonalization.}
%%\vspace*{-0.5cm}
\label{fig:green-park}
\end{figure}

Our preliminary analysis of both datasets shows that aggregate time series of the ROIs (tube stations or regions) exhibit no particular trend but do preserve weekly/daily seasonality. Therefore, as proposed in prior work, e.g.,~\cite{hylleberg2014seasonality}, we de-seasonalize each region's time series via additive decomposition. More specifically: %%\vspace*{-0.15cm}
\begin{equation}
D_{it} = Y_{it} - \bar{Y_{it}}%%\vspace*{-0.1cm}
\end{equation}
where $Y_{it}$ is the ROI's original time series $i$ and $\bar{Y_{it}}$ is its seasonality defined as
$\bar{Y_{it}} = { \frac{1}{w} \sum Y_{idh} }$,
with $w$ being the number of weeks in the dataset, $d$ the day of the week (i.e.,  $d \in$ $ \{\text{Monday}, \ldots, \text{Sunday}\}$), and $h \in \{0, \ldots, l\}$ with $l$ denoting the number of epochs in one day (24 since we aggregate hourly). Observe that $\bar{Y_{it}}$ is a time series containing the average value of each specific time slot (e.g., Mondays 3pm -- 4pm). As an example, Figure~\ref{fig:green-park} shows Green Park station's (a station among the busiest TFL stations) aggregate time series in both its original and de-seasonalized form. Note a negative spike on the morning hours of March 8, as the station must have probably had reduced access (e.g. due to partial closure).  
In general, the de-seasonalized ROIs' time series show strong auto-regressive structure.

\subsection{Predicting Traffic Volumes in ROIs}
\label{sec:predictions}

We now investigate how to make hourly traffic volume predictions on ROIs using the TFL and SFC time series. Such predictions are particularly useful in modern cities for journey planning~\cite{lam2005short,stathopoulos2003multivariate}, congestion prevention~\cite{silva2015predicting} as well as improving transportation service levels and adjusting staff needs at stations~\cite{tfl-budget}. 

We focus on the 100 busiest TFL stations and the 100 most popular SFC regions. Since our preliminary analysis shows that ROIs' time series are stationary and exhibit strong auto-regressive structure, we turn to ARMA modeling (cf.~Section~\ref{sec:arma}). For each ROI $i$, we feed the ARMA model with the values of the last 6 days of its aggregate and de-seasonalized time series $Y_{it}, D_{it}$, respectively. We train the model using the first 5 days of $D_{it}$ and test it against the last (``test day'') of $Y_{it}$ following a recursive approach with a sliding time window to predict its hourly traffic. To do so, for each time slot we combine ARMA model's predictions on $D_{it}$ with ROI's seasonality $\bar{Y_{it}}$, therefore, our predictions are given by
$\hat{Y_{it}} = \hat{D_{it}} + \bar{Y_{it}}$,
where $\hat{D_{it}}$ is the ARMA prediction on the ROI's de-seasonalized time series and $\bar{Y_{it}}$ its seasonality. 

\begin{figure}[t]
\centering
\begin{subfigure}[b]{0.355\textwidth}
   \includegraphics[width=1\linewidth]{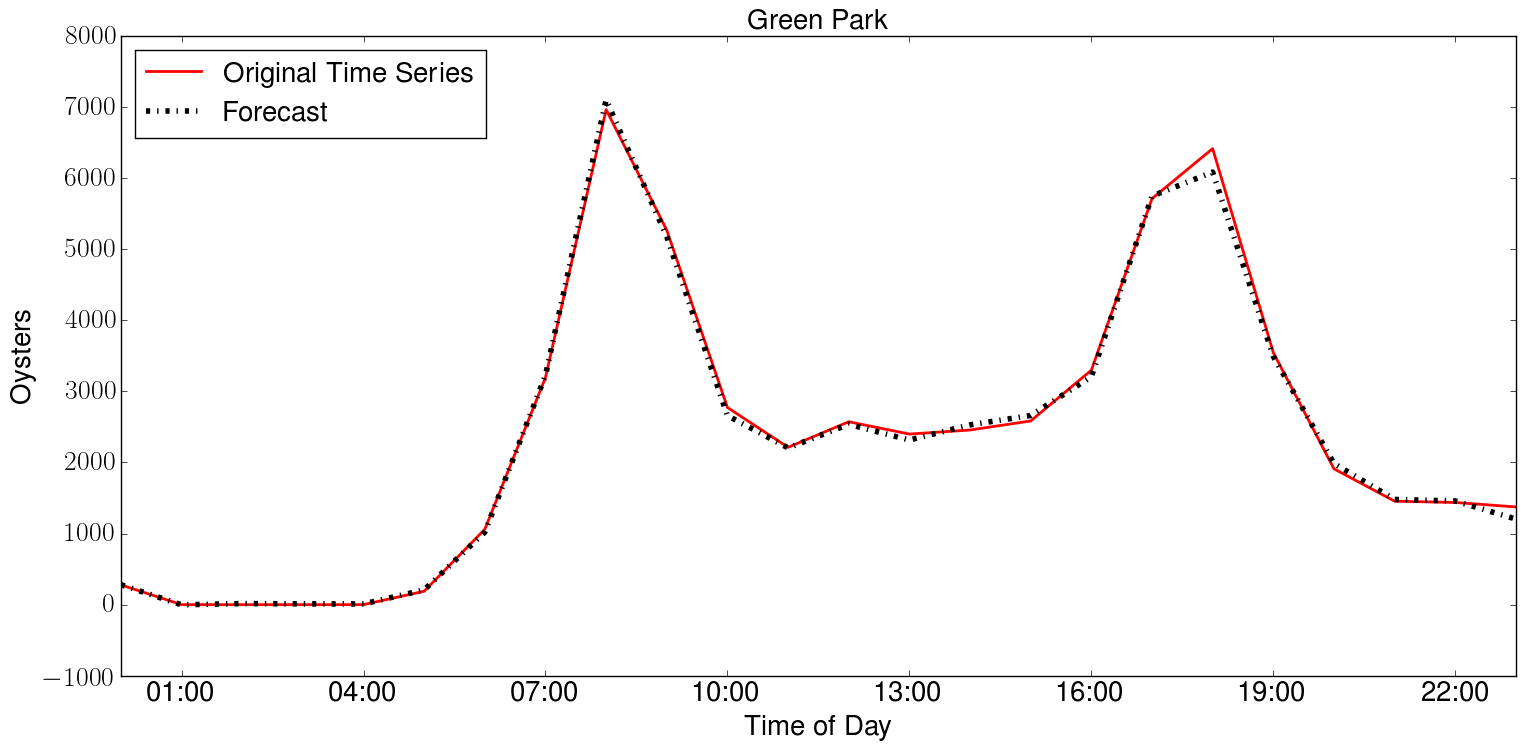}
   \caption{}
   \label{fig:green-park-des-fit-combined} 
\end{subfigure}
\begin{subfigure}[b]{0.355\textwidth}
   \includegraphics[width=1\linewidth]{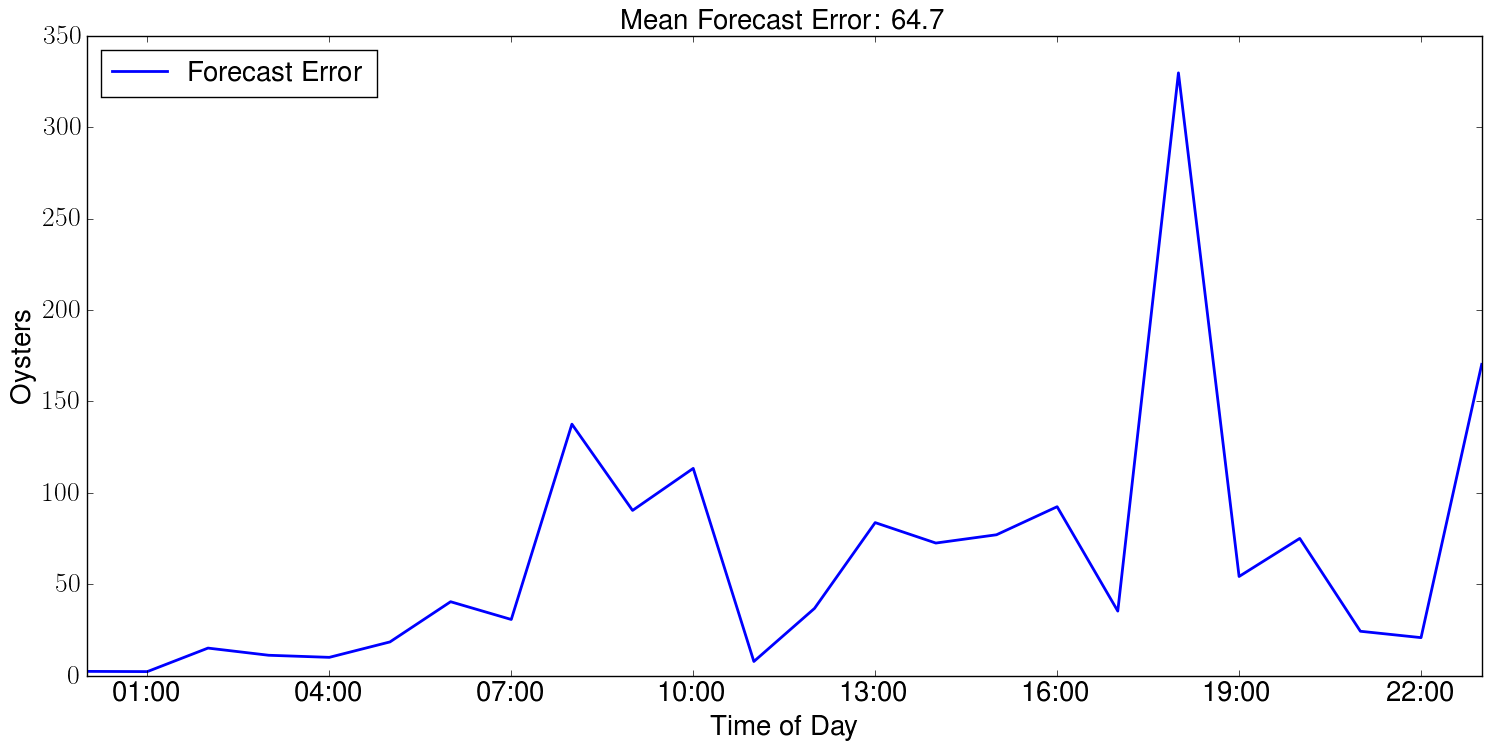}
   \caption{}
   \label{fig:green-park-des-err}
\end{subfigure}
\caption{Hourly traffic forecasts for Green Park station on March 25.}
\label{fig:green-park-des-forecast}
 %%\vspace*{-0.3cm}
\end{figure}

We then compare our approach against a baseline, i.e., a black box approach where we fit the ARMA model directly on each ROI's $i$ time series $Y_{it}$. We evaluate the accuracy of the predictions using the absolute forecast error as $
e_{it} = \mid Y_{it} - \hat{Y_{it}}\mid $,
where $Y_{it}$ is the actual time series value at time slot $t$ (ground truth) and $\hat{Y_{it}}$ is the predicted value for that time slot using our approach. We also convert the error into a percentage error, i.e.,
$p_{it} = \frac{e_{it}}{Y_{it}} \times 100$.
Figure~\ref{fig:green-park-des-fit-combined} plots the traffic volume forecast for Green Park station on March 25, while Figure~\ref{fig:green-park-des-err} shows the absolute forecast error. Overall, on the TFL dataset, the mean absolute forecast error for March 25, over the 100 busiest stations, is $59.53 \pm 42.48$ oysters,
compared to $545.9 \pm 376.8$ oysters with the baseline. This corresponds to an error of $19.6\% \pm 59.5\%$ vs  
$638\% \pm 1619\%$, showing that the seasonality-based method significantly outperforms the baseline.

\begin{figure}[t]
\centering
\begin{subfigure}[b]{0.355\textwidth}
   \includegraphics[width=1\linewidth]{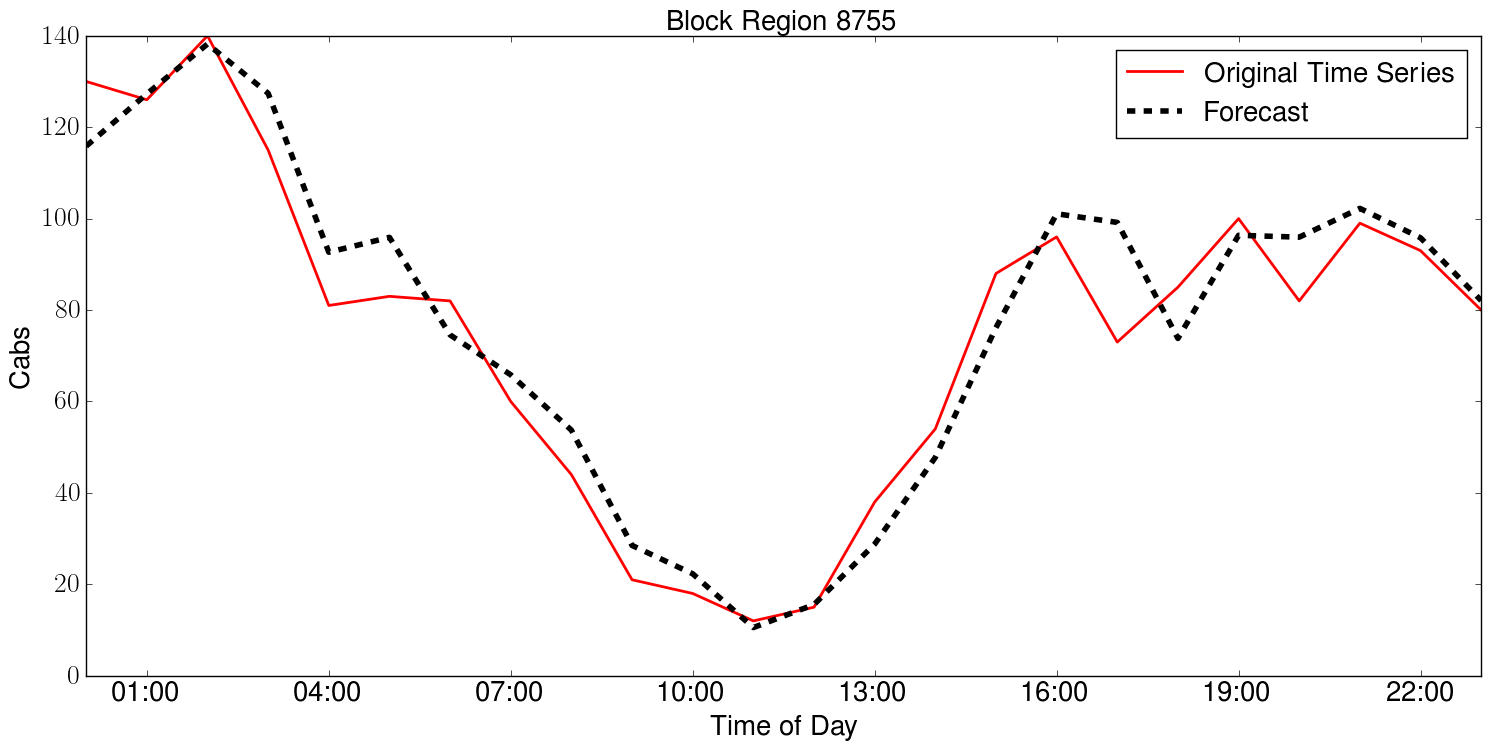}
   \caption{}
   \label{fig:region-8755-preds} 
\end{subfigure}
\begin{subfigure}[b]{0.355\textwidth}
   \includegraphics[width=1\linewidth]{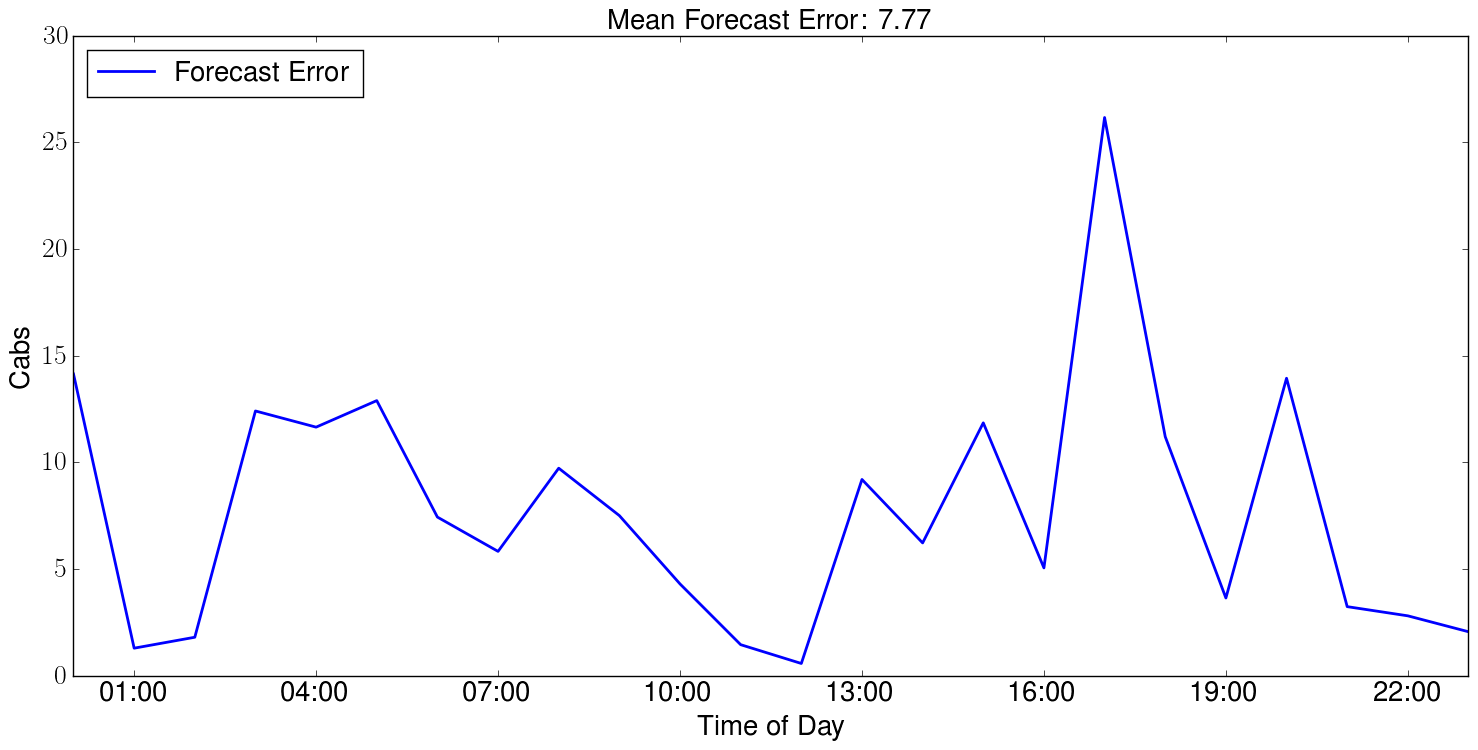}
   \caption{}
   \label{fig:region-8755-for-err} 
\end{subfigure}
\caption{Hourly traffic forecasts for SFC's region 8755 on June 5.}
\label{fig:region-8755-preds-err}
 %%\vspace*{-0.3cm}
\end{figure}

We follow the same approach for the SFC dataset, predicting the traffic volume of the most popular regions. Figure~\ref{fig:region-8755-preds-err} shows the  predictions and the forecast error for the region with identifier 8755, on June 5. The average forecast error over the 100 busiest regions is $5.62 \pm 3.12$ taxis ($19.7\% \pm 10.3\%$), whereas, the baseline error is $9.07 \pm 2.95$ ($35.4\% \pm 13.4\%$), once again showing that predictions can be improved when considering seasonal effects.

\subsection{Detecting Traffic Anomalies}\label{sec:anoms}
Next, we focus on detecting traffic volume anomalies on ROIs' time series. This is particularly important for traffic provisioning and travel planning as trip recommendations can be made to drivers/commuters during road accidents, incidents or events that cause overcrowding in transportation stations~\cite{tfl-twitter,waze,pan2013crowd}.

Once again, we utilize ARMA modeling: our intuition is to train the model for each ROI and rely on the absolute forecast error as a confidence interval for detecting anomalies. More precisely, since the forecast error is normally distributed, we apply the $3\sigma$ rule and set upper and lower confidence intervals $\lambda_{1}, \lambda_{2}$. Thus, we detect an anomaly when: %
$e_{t} > \lambda_{1} \parallel e_{t} < \lambda_{2}$,
where $\parallel$ is the logical OR operator, $\lambda_{1} = \mu + 3 \sigma$, $\lambda_{2} = \mu - 3 \sigma$, and $\mu, \sigma$ are, respectively, average and standard deviation of the forecast error $e_t$. In a way, we flag as anomalies time slots that our model could not predict with good accuracy. 

Subsequently, we experiment with our anomaly detection technique using a similar approach to that described in Section~\ref{sec:predictions}: we train the ARMA model using data of the first week of each ROI's time series and test it against the remaining weeks (i.e., for the TFL dataset we have 3 test weeks while for the SFC we have 2), using a sliding window, aiming to identify traffic volume anomalies. We focus on the 100 busiest stations of the TFL dataset and detect $896$ anomalies, which roughly corresponds to $2\%$ of all 1-hour slots in the 3 test weeks. In the SFC dataset, over the 100 busiest regions, we find $366$ anomalies (i.e., $1\%$ of the 2 test week time slots). %
We rank each anomaly based on its deviation from the confidence intervals as a measure of its magnitude and keep track of the top $10\%$ of anomalies, i.e., 90 anomalies for the TFL dataset and 30 for SFC: in Section~\ref{sec:improv}, we will investigate whether or not we can enhance traffic prediction in the presence of anomalies by combining information from correlated ROIs.

Note that we do not have \textit{ground truth} as to what constitutes an actual {\em ``anomaly''} in our datasets, so we cannot empirically evaluate how well our approach corresponds to detecting, e.g., events, strikes, disruptions, etc., and anyway this would be out of the scope of our work. In general, we consider an anomaly to be a pattern that does not conform to expected normal behavior~\cite{chandola2009anomaly} and, as such, our anomaly detection techniques really consist in automatically \textit{flagging} such patterns using aggregate locations. %
By focusing on the top events in terms of deviation from the confidence intervals (i.e., unexpectedly increased/decreased traffic patterns in ROIs), we aim to investigate whether collecting information from multiple ROIs can improve traffic volume predictions in the presence of anomalies (see Section~\ref{sec:improv}). 

\begin{figure}[t]
\centering
\begin{subfigure}[b]{0.355\textwidth}
   \includegraphics[width=1.0\linewidth]{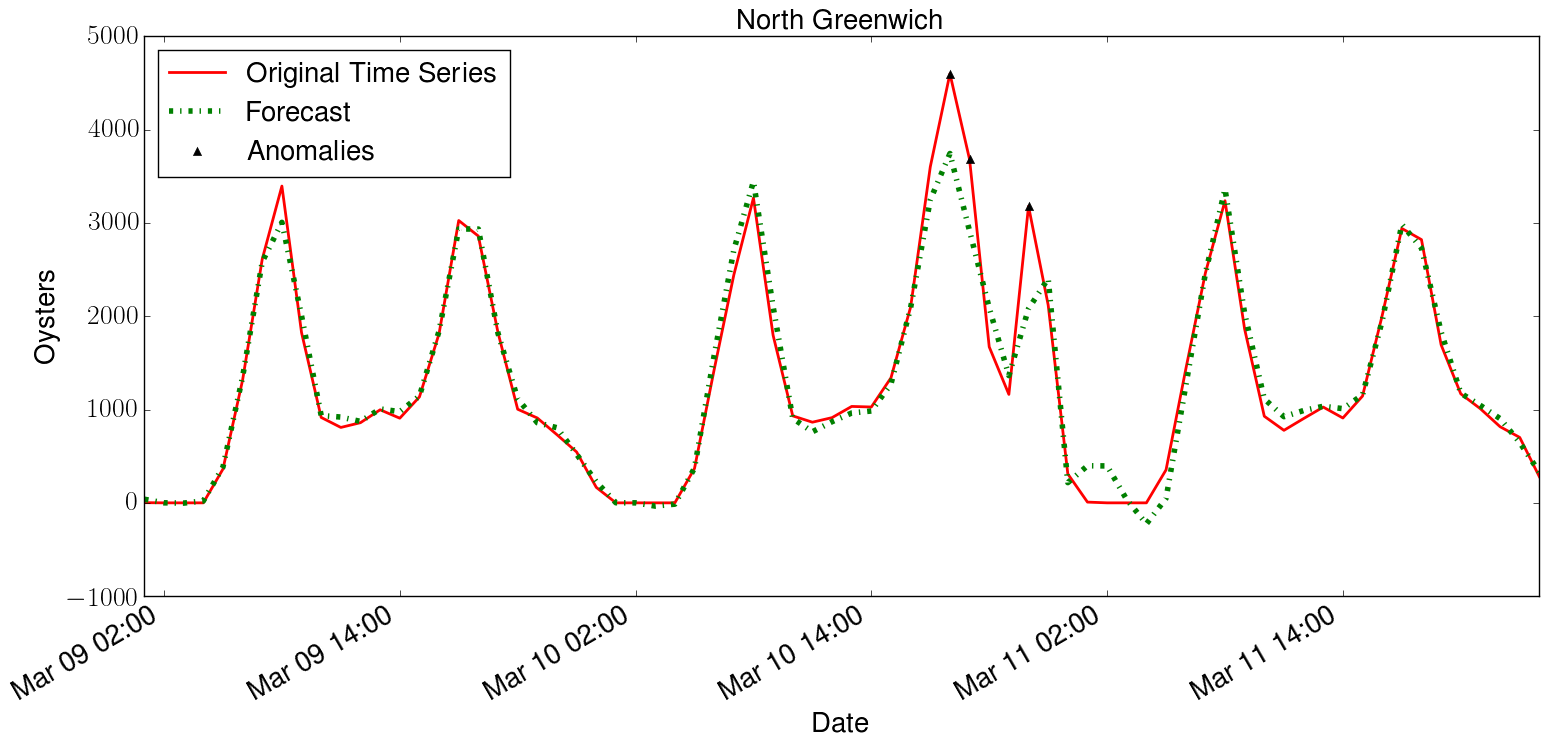}
   \caption{}  
   \label{fig:north-green-anom}      
\end{subfigure}
\begin{subfigure}[b]{0.355\textwidth}
   \includegraphics[width=1.0\linewidth]{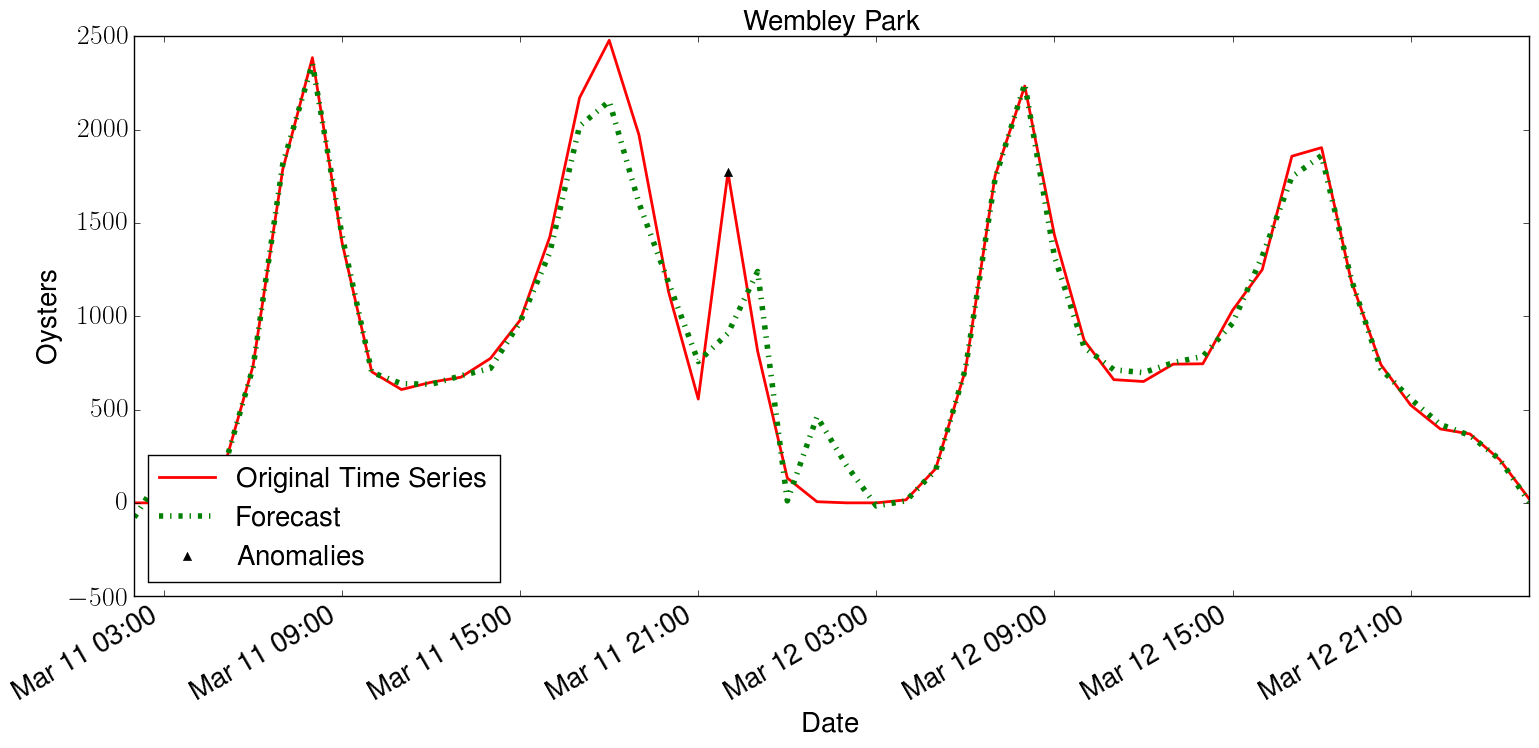}
   \caption{}  
   \label{fig:wembley-anom}
\end{subfigure}
\caption{Original time series, forecasts, and detected anomalies in (a) North Greenwich station, March 9--11, and (b) Wembley Park station, March 11--12.}
 %%\vspace*{-0.4cm}
\end{figure}

\begin{figure}[t]
\centering
\begin{subfigure}[b]{0.355\textwidth}
   \includegraphics[width=1.0\linewidth]{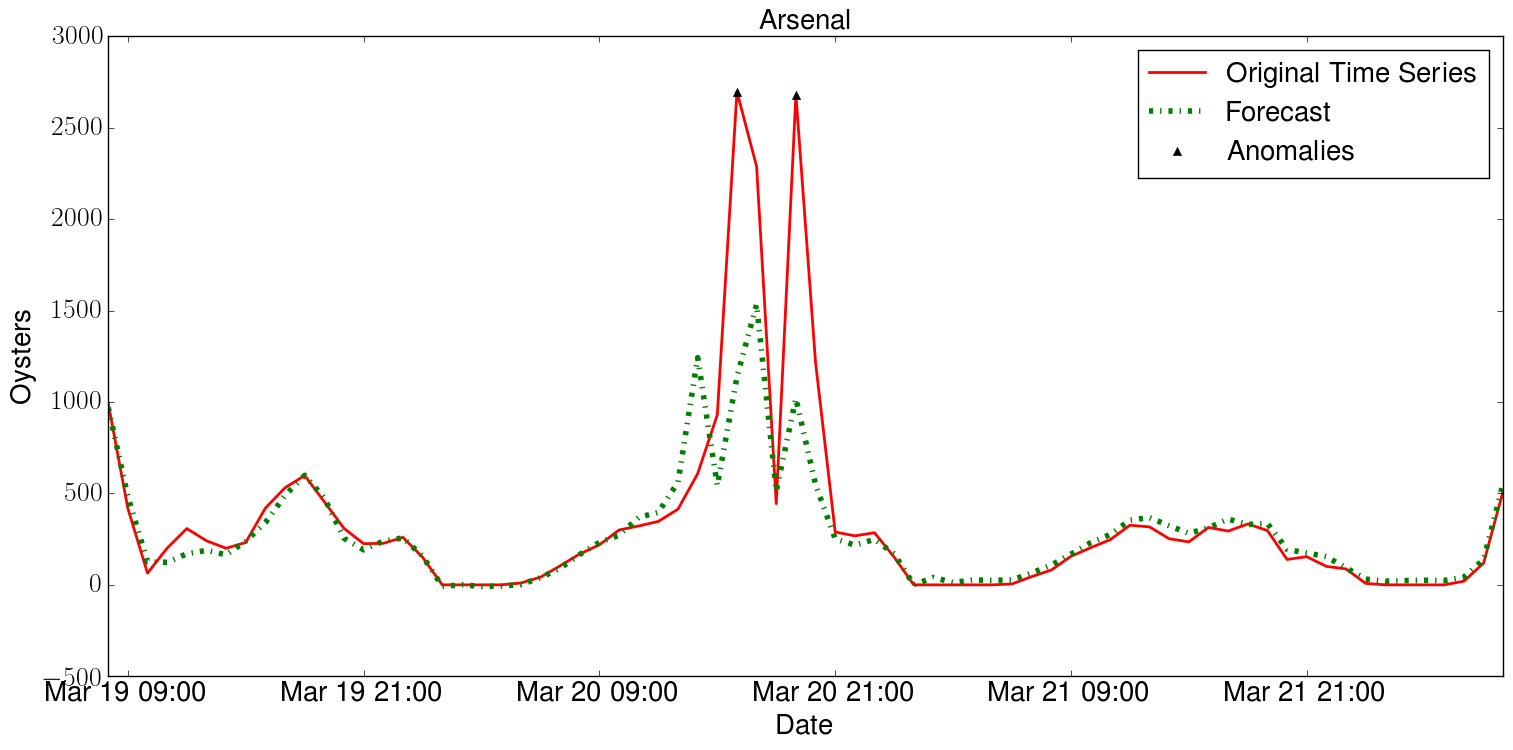}
   \caption{}  
   \label{fig:arsenal-anom}      
\end{subfigure}      
\begin{subfigure}[b]{0.355\textwidth}
   \includegraphics[width=1.0\linewidth]{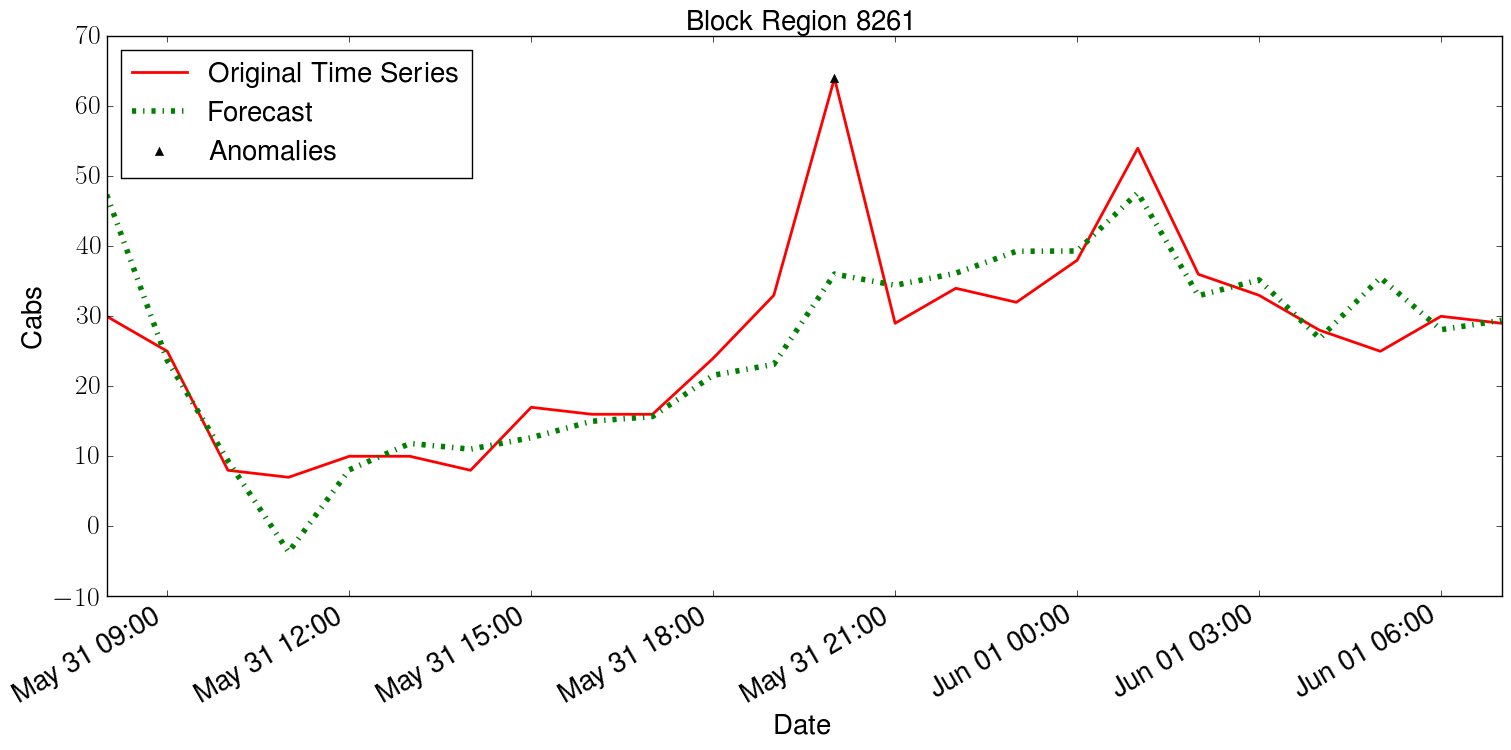}
   \caption{}
   \label{fig:region-8261-ts-anoms} 
\end{subfigure}
 \caption{Original time series, forecasts, and detected anomalies in (a) Arsenal station, March 19--22, and (b) SFC region 8261, May 31--June 1.}  
 %%\vspace*{-0.5cm}
\end{figure}

We discuss some case studies among the top anomalies that we were able to correlate with external events. %
As shown in previous work~\cite{sagl2012visual}, distinct human mobility patterns are observed during events that attract big crowds like football matches or music concerts. In Figures~\ref{fig:north-green-anom} and~\ref{fig:wembley-anom}, we plot ``anomalies'' we detect in North Greenwich and Wembley Park tube stations, on the evening hours of March 10 and March 11. This seems to correspond to concerts taking place in the O2 and Wembley arenas, which are venues close to those stations. These events likely cause increased traffic spikes at nearby stations. Similarly, Figure~\ref{fig:arsenal-anom} shows the original aggregate time series of Arsenal station as well as the anomalies detected on it when fitting our model. We can observe that the model detects anomalies on the evening of March 20, when an Arsenal FC soccer game was taking place. Finally, Figure~\ref{fig:region-8261-ts-anoms} does the same for a region (id 8261) in the SFC dataset that is nearby AT\&T Park, showing increased taxi traffic on the evening of May 31, when the San Francisco Giants were playing a baseball match.

\subsection{Predicting Traffic Volumes in Case of Anomalies}\label{sec:improv}
We now investigate how to improve ROI traffic volume predictions in the presence of anomalies, using additional information from correlated ROIs. To this end, for each ROI, we use Spearman correlation (see Section~\ref{sec:prelim}) in order to discover those ROIs whose traffic can be useful for enhancing our predictions. Subsequently, we train a VAR model -- geared to capture linear dependencies among multiple time series -- with the time series of a ROI as well as the time series of its correlated ROIs and we compare the prediction results against a \textit{local} model, i.e., an ARMA model trained only with ROI's past local information (note that the ARMA model described in Sections~\ref{sec:predictions} and~\ref{sec:anoms} now consists our \textit{baseline}). 
In the rest of this section, for each of our datasets, we start by describing our approach on a specific case study and then we generalize our results by focusing on the top events (90 for TFL and 30 for SFC, respectively) that our anomaly detection module has flagged as possible anomalies. 

\descr{TFL.} We first focus on Saturday March 20, when our anomaly detection module spots anomalies -- i.e., increased traffic volume -- on the Arsenal station, likely caused by an Arsenal FC soccer game. We zoom in on the two hours before and after the game (15:00--17:00, resp., 19:00--21:00, respectively) when the majority of Arsenal fans, exit from, resp., enter Arsenal station. We follow a similar aggregation approach as that described in Section~\ref{sec:datasets} although now %
for each station $i$, we keep two separate time series: one counting passengers entering the station ($Y^{in}_{it}$) and one counting those exiting it ($Y^{out}_{it}$). Once again, we de-seasonalize each station's entering/exiting time series as discussed in Section~\ref{sec:seasonal}.

To discover stations correlated with Arsenal, we compute the Spearman correlation (see Section~\ref{sec:prelim}) between the de-seasonalized time series of passengers entering/exiting Arsenal as well as the de-seasonalized time series of all the remaining stations in the TFL network, by sliding them up to 1 hour earlier/later. Our results show that the traffic exiting at Arsenal is highly correlated with the traffic entering at various other TFL stations including Arnos Grove, King's Cross, Leicester Square, Blackhorse Road and Cockfosters (i.e., stations on the same line with Arsenal or on a line connected with the line of Arsenal). 
We then set to improve the traffic volume predictions of passengers exiting at Arsenal station before the match, by feeding a model with the de-seasonalized entering time series of the correlated stations. To do so, we use a vector autoregression model (VAR) (see Section~\ref{sec:prelim})  which describes the evolution of a set of variables over the same sample period as a linear function of their past values.
\begin{figure}[t]
\centering
   \includegraphics[width=0.42\textwidth]{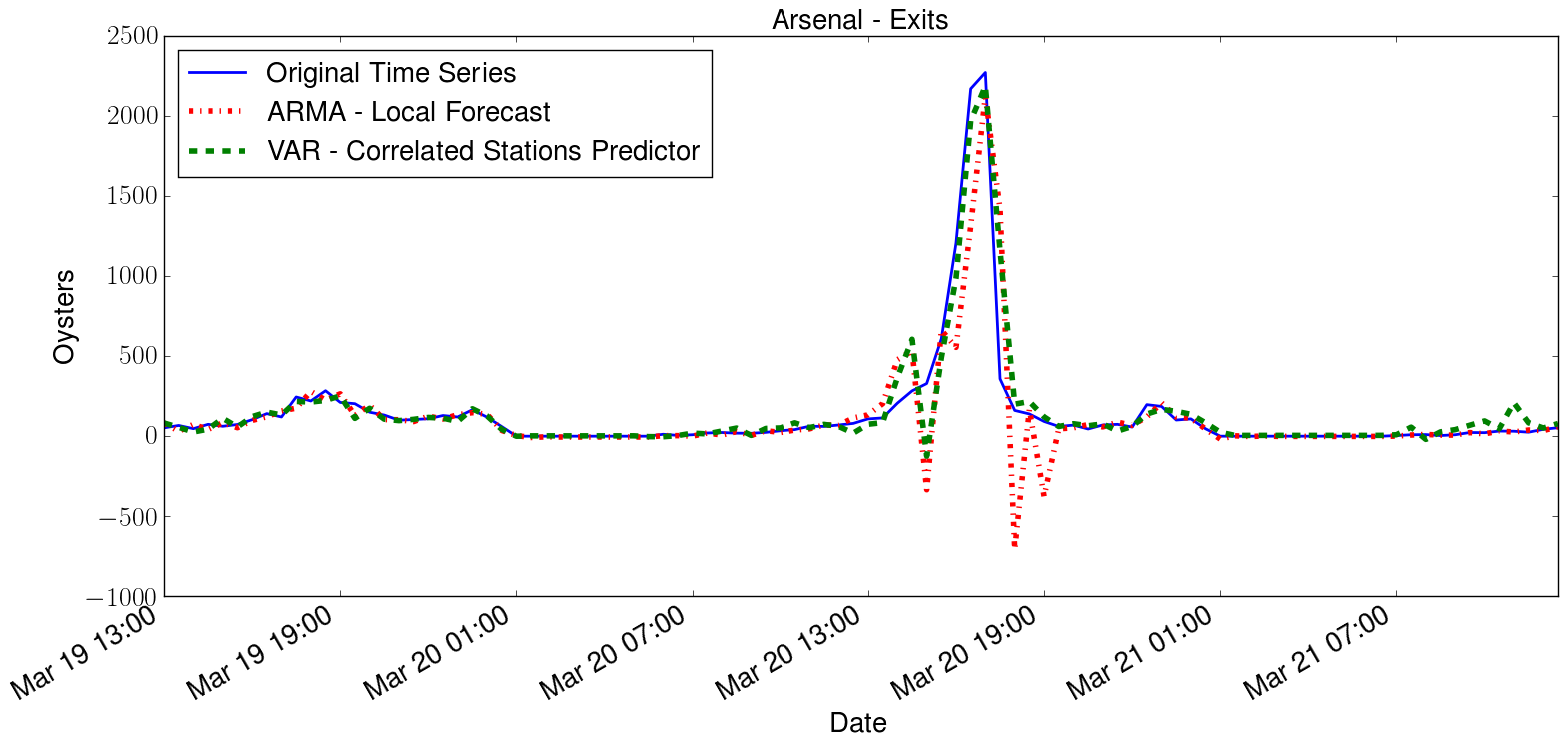}
   \caption{Forecast of passengers exiting Arsenal station March 19--21 based on a local ARMA model only vs. the VAR model with information from correlated stations.}
   \label{fig:arsenal-match-day} 
%%\vspace*{-0.2cm}
\end{figure}
Figure~\ref{fig:arsenal-match-day} shows traffic volume predictions for passengers exiting Arsenal station using the ARMA model with local information (i.e., the station's past exiting time series, which now consists our \textit{baseline}) and using the VAR model enhanced with additional information from the 10 most correlated stations. We observe that the \textit{enhanced} model makes significantly better predictions between 15:00--17:00 on March 20, where we observe increased traffic due to the game. We measure the average forecast error of the exit traffic predictions of Arsenal on that day as $133.9 \pm 270.6$ oysters (i.e. $93\% \pm 185\%$) when making predictions with local information only and $65.26 \pm 135.04$ oysters (or $59.1\% \pm 57.4\%$) when enhancing it with the correlated stations. %

In relation to the same event, we discover high correlation between the time series of passengers entering at Arsenal and those exiting at Arnos Grove a time slot later. Thus, we enhance the local ARMA prediction of traffic exiting volume at Arnos Grove by feeding the model with the de-seasonalized traffic entering Arsenal. Again, we use VAR to calculate the appropriate values for the parameters.
The enhanced model makes  better predictions between 19:00 and 21:00 where there was increased exiting traffic due to the Arsenal game. Indeed, the average forecast error for March 20 decreases from $23.38 \pm 48.79$ oysters to $11.22 \pm 18.61$, a $52\%$ average improvement in traffic volume predictions. 

Next, we apply our approach for {\bf\em all the top 90 anomalies} found in the TFL dataset (cf.~Section~\ref{sec:anoms}), i.e., for each station under the presence of an anomaly, we predict its exiting or entering traffic (depending on which direction the anomaly has been detected) using a VAR model trained with the exiting or entering time series (again, depending on the anomaly direction) from $10$ correlated stations. We measure the average forecast error over the day of the anomaly and we compare it against a local approach (ARMA model), where predictions are made using only past station's information. Overall, for the 90 anomalies of the TFL dataset, we observe a $29\% \pm 13\%$ improvement in our predictions when employing the VAR model, indicating that sharing information between correlated ROIs improves prediction quality.

\descr{SFC.} We follow a similar approach for anomalies detected on the SFC dataset. Our correlation analysis shows that the de-seasonalized time series of neighboring regions have high correlation, as it is likely that they are connected by the same roads. For instance, if we focus on the anomaly detected in region 8556 between May 27--28, we observe a $41\%$ improvement in predictions when training a VAR model including additional information from $5$ correlated regions (i.e., block regions with ids 8557, 8657, 8558, 8655 and 8555) %
compared to the baseline, i.e., the ARMA model that predicts using only local information. 
Similar to our TFL experiments, we generalize this approach by trying to improve the predictions for the {\bf\em top 30 anomalies} of the SFC dataset, enhancing our model with information from $5$ correlated regions. In this case, 
we obtain a $18\% \pm 14\%$ average improvement on the prediction.

\descr{Discussion.} We observe that our techniques discussed above yield better improvements over the TFL dataset compared to SFC. A possible explanation arises from the nature of the TFL dataset and the way passenger trips are aggregated, i.e., at station level while preserving the notion of \textit{direction} (number of passengers exiting or entering a station). This allows us to perform a more fine-grained correlation analysis in comparison to the SFC dataset, where as we aggregate the GPS locations, we lose the granularity of each taxi's trajectories (i.e., taxi moving from one region to another). While this is a good feature vis-a-vis privacy guarantees, it motivates the need to gather (privacy-friendly) aggregate location statistics while preserving directions.
%

%%\vspace*{-0.1cm}
\section{A System for Privacy-Friendly Mobility Analytics}
\label{sec:system}
After having assessed the usefulness of collecting and using aggregate locations for mobility analytics, we now set to investigate how to {\em enable} such collection in a privacy-friendly way.
To this end, we design 
a distributed, collaborative framework whereby users install an application -- called {\em Mobility Data Donors} (MDD) --
that regularly monitors their locations, stores it locally, and periodically reports it to our server in a privacy-friendly way.\footnote{\small The app Android prototype  is available upon request.} Privacy is guaranteed through aggregation, by means of the scalable private aggregation protocol 
presented in Section~\ref{sec:aggr}, thus, the server only learns aggregate information, i.e., how many (but not which) users were in a particular region or entered/exited a particular underground station in an interval of time.
Once the server has received the aggregate location data (i.e., counts of users' presence in ROIs), it can use it for mobility analytics applications. 

As discussed earlier, protecting privacy of user locations is critical, 
as sensitive data about individuals, such as their religion~\cite{taxis-rainbows}
or their identity~\cite{golle2009anonymity} can be inferred, and even a few locations are enough to re-identify users
from anonymized traces~\cite{zang2011anonymization}.
The ability to privately collect location reports enables applications that would otherwise be impossible due to privacy concerns. For instance, obtaining data from
TFL typically requires several rounds of NDAs and the promise not to re-distribute the data: although TFL could publish aggregate statistics, it is unlikely they would do so in real-time (a crucial aspect for mobility analytics) and anyway this would only capture one aspect of urban mobility---i.e., underground/overground trips but not, e.g., taxis or buses. In general, collecting locations directly from the users, without requiring them to forego their privacy, paves the way for a number of novel and interesting analytics, which we are confident our work will support.

\subsection{Data Collection}
To support private data collection, we need a secure aggregation protocol that allows a server to only learn aggregate locations. 
As discussed in Section~\ref{sec:aggr}, we choose the one by Melis et al.~\cite{melis} as it supports scalability and fault-tolerance without the need for a trusted third party, which are fundamental factors for the success of a distributed, crowd-sourcing system. 
Our system model mirrors to that of Melis et al.~\cite{melis}, i.e., it consists of a server, or aggregator, that  facilitates networking and collects aggregate location counts from a set of mobile users running the MDD app. There is no other trusted authority. As in~\cite{melis}, we assume the aggregator and the users to be honest-but-curious, i.e., they follow protocol specifications and do not misrepresent their inputs, but try to extract information from other parties.
When installed, the MDD app generates a private/public key pair (see setup phase in Section~\ref{sec:aggr}) and communicates its public part to the aggregator. After setup, the app runs on the background, regularly collecting GPS coordinates. At predefined time slots (by default, every hour), the privacy-preserving aggregation is triggered by the server, provided that there are at least $\tau$ users connected, which are randomly assigned to groups of $u$ users. In the default setting,
the app maps GPS coordinates to a grid of $p \times p$ cells of $\rho$ square miles, and
builds a $p \times p$ matrix corresponding to the grid,  setting to 1 items corresponding  
to ROIs the user has visited in the specified time slot (and 0 otherwise).\footnote{\small Note that the app is easily tunable so that, instead of binary values, the matrix encodes, e.g., duration of user's presence in each ROI, whether the user has entered or exited a cell, etc.}
The values of $u$, $\tau$, and $\rho$ are passed onto the user to inform them of the granularity of the data collection, and give them the option to withdraw (minimum acceptable values can be adjusted from the MDD's settings).
Next, as per~\cite{melis}, the app generates blinding factors (summing up to zero) based on the keys of the users in the same group, and encrypts each entry in the matrix. Finally, it sends the encrypted matrix to the aggregator who obliviously aggregates all (encrypted) matrices and decrypts the aggregate location counts used for the analytic tasks.

Besides recording coordinates, and mapping them onto a grid, the app can also recognize points of interest, such as train/underground stations, which is particularly useful for  mobility analytics on transport datasets such as the TFL data. In this case, the aggregation takes place on a vector where each item corresponds to a point of interest and is set to $1$ if the user has visited it in the specified time slot. 
\subsection{Experimental Evaluation}
Next, aiming to assess the real-world deployability of our techniques, we empirically evaluate the performances of the MDD app, in terms of computation, communication, and energy overhead. Specifically, we evaluate the overhead imposed by the cryptographic operations needed for the privacy-preserving data collection. We use the mobility datasets from Section~\ref{sec:datasets} as guidelines for simulating the system. For our experiments, we use the prototype implementation, in Javascript/Node.js, of the protocol by Melis et al.~\cite{melis} and have adapted its client-side to run on Android using Apache Cordova.\footnote{\small{\url{https://cordova.apache.org/}}} The cryptographic operations are implemented using elliptic curve cryptography, specifically, the Ed25519 elliptic curve~\cite{bernstein2012high} (supporting 256-bit points and offering 128-bit security) from the Elliptic.js library.\footnote{\small{\url{https://github.com/indutny/elliptic}}} 

For the sake of our evaluation, we run the experiments on a mid-range (rather than a high-end)
Android device, as we do not want to limit deployment only to (possibly higher-income) users that have newer phones.
We use a Samsung Galaxy A3 device, equipped with a 1.2 GHz quad-core Snapdragon 410 processor and 1.5GB RAM, running Lollipop v5.0.2. For our energy consumption analysis, we utilize PowerTutor~\cite{power-tutor}, an Android app for power monitoring.
Note that, although a Javascript implementation of the cryptographic operations might not be optimal in terms of efficiency (e.g., compared to Java), it offers portability among different mobile OSes. Anyway, we have actually benchmarked a Java implementation of the same operations and obtained similar results.

\begin{figure}[t]
\centering
   \begin{subfigure}{0.355\textwidth}
   \includegraphics[width=1\textwidth]{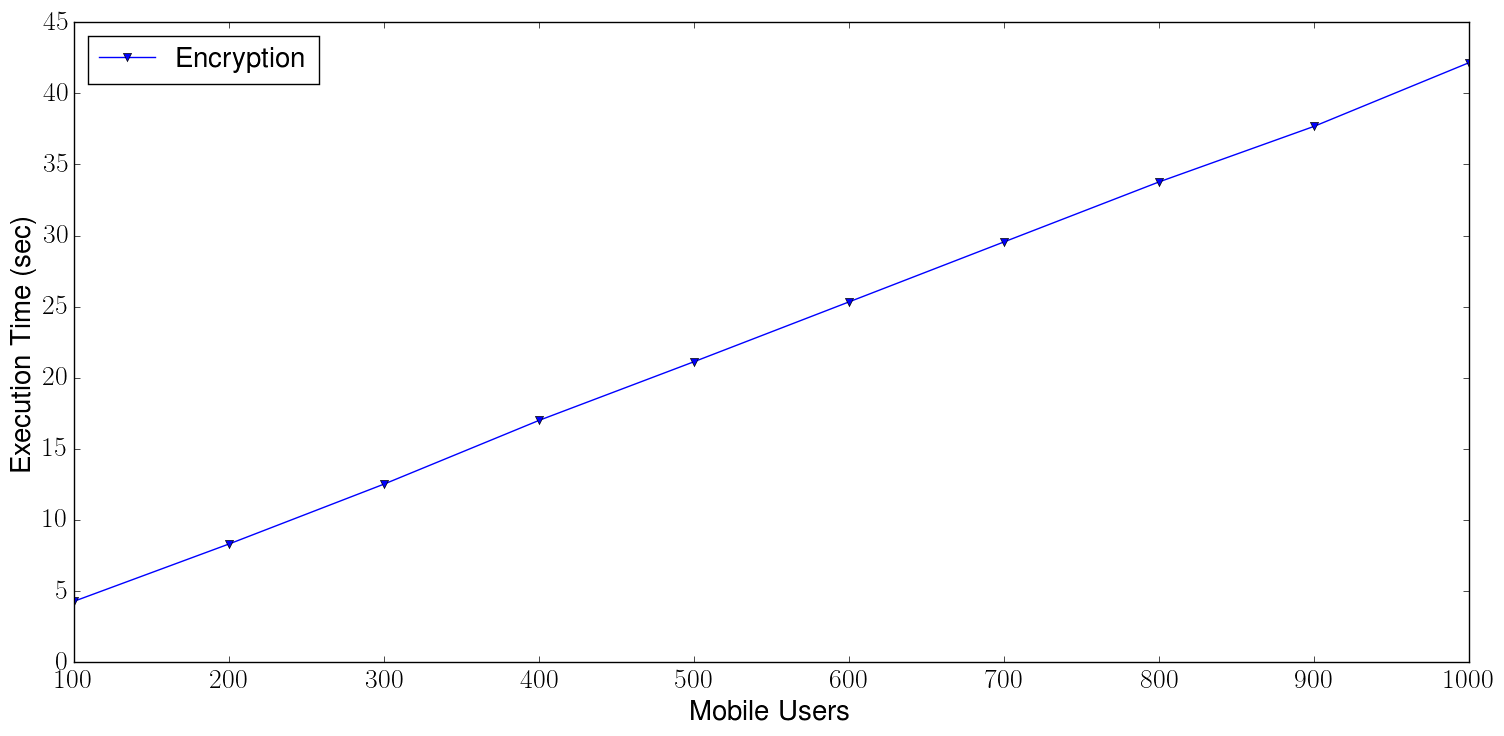}
   \caption{TFL}
   \label{fig:tfl-execution-time} 
\end{subfigure}
\begin{subfigure}{0.355\textwidth}
   \includegraphics[width=1\textwidth]{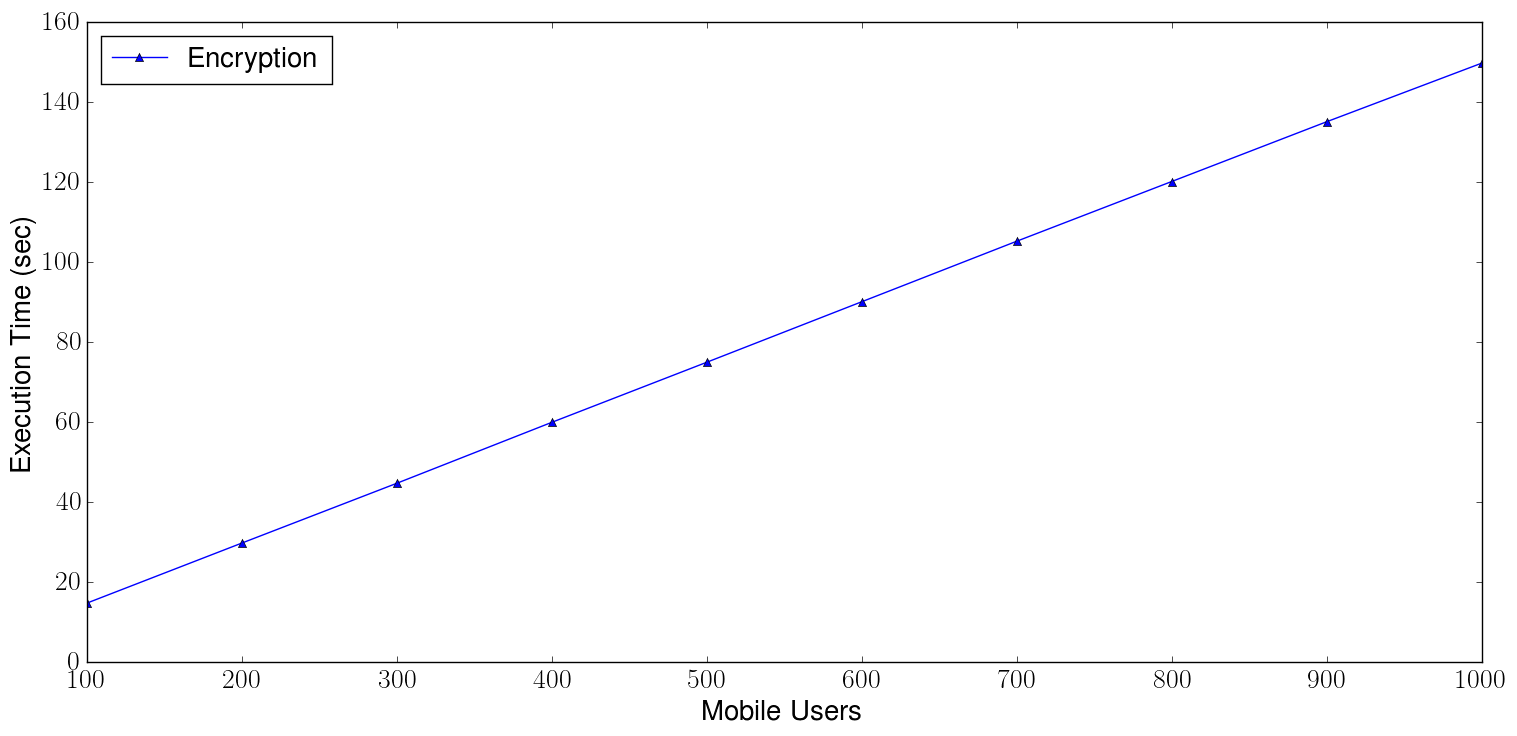}
   \caption{SFC}
   \label{fig:sfc-execution-time} 
\end{subfigure}
\caption{Execution times of the encryption phase for increasing number of users in (a) TFL and (b) SFC settings.}
 %%\vspace*{-0.4cm}
\end{figure}

\descr{TFL.} We start our experiments with the TFL use-case. Recall that the TFL data involves $582$ ROIs (stations), so each user device, for each time slot $t$, encrypts a matrix of size $2 \cdot 582$, with the first row indicating entering the station and the second exiting it.
We here remind that the complexity of the aggregation protocol depends on how many users are assigned to the same group, since the blinding factors are derived from public keys of other users in the group.
In Figure~\ref{fig:tfl-execution-time}, we plot the execution time, measured on our Android device, of the encryption phase vis-a-vis the number of users in the group. As expected, running times grow linearly in the size of the group. For instance, the encryption performed by each mobile device takes $4.2s$ with 100 mobile users and $42s$ with 1,000. Therefore, one should 
probably keep groups at around 200 users, which offers a reasonable trade-off between granularity (in terms of privacy) and efficiency. Obviously, even if, say, 1 million London commuters were to participate, the system simply scales by running multiple parallel instances with each group, and combining multiple aggregates from 5,000 200-user groups.
 
When assigning users to groups of size 200, in each round of the protocol, each device has to download $10.7KB$ worth of public keys, and transmitting encrypted values for $2 \cdot 582$ entries results in an overhead of $4.54KB$ (independently of how many users are in the group). Finally, we measure the energy consumption to be $862mJ$ for the encryption part, $609mJ$ to download public keys via Wi-Fi, and $322mJ$ to transmit the encrypted matrix. We observe that the energy overhead is quite small for a modern mobile phone (for reference we note that downloading a $20KB$ web page with Google Chrome via WiFi consumes approximately $800mJ$).

\descr{SFC.} Next, we evaluate complexities considering the SFC use-case, for which we divide the city of San Francisco into a grid of $100 \times 100$ cells. In this setting, our mobile app, for each time slot $t$, encrypts a matrix of size 10,000. Figure~\ref{fig:sfc-execution-time} displays the resulting computational overhead for an increasing number of users. Once again, we observe execution times linear in the number of users in the system, i.e., the encryption phase requires about $14s$ with 100  and $149s$ with 1,000 users. We also measure communication and energy overhead, assuming groups of 100 users.
Obtaining the public keys of users in the group requires downloading $5.37KB$, while transferring the encrypted $100 \times 100$ matrix requires $39KB$, which, once again, is acceptable for a mobile app. The energy consumption of the cryptographic operations in each protocol round is $485mJ$. Finally, the communication operations (via WiFi) require $306mJ$ and $2769mJ$, respectively.

\descr{Succinct Data Representation.} We observe that the number of users in each group mainly affects the computation overhead, while the communication and energy overheads are primarily influenced by the size of the input. 
In particular, from our experiments above, we notice that the computation/communication/energy is appreciably low
when groups are in the order of hundreds of users and matrices are in the order of thousands. This means that if aggregation is performed over larger inputs, the protocol would quickly incur high energy, communication, and computation overheads, and this would remarkably limit the deployability of our techniques. 

In fact, even if in the use-cases considered in this paper the overhead is appreciably low, 
we would not be able to extend to, e.g., building origin-destination matrices~\cite{silva2015predicting,zhong2016variability,munizaga2012estimation},
rather than only keeping counts. Origin-destination information is particularly useful to obtain finer grained statistics, e.g., discovering \textit{similar} locations for personalized recommendations~\cite{clements2011personalised}, and/or modeling the effects of disruptions in a transportation network as in~\cite{silva2015predicting}. 
More specifically, an origin-destination matrix for the TFL data would be of size $582 \times 582$. At each time slot, the app sets element (i, j) in the matrix to 1 if the user commuted from station i to station j. For groups of 100 users, this would result in a $7.6$s computation overhead (with $3$,$387mJ$ energy overhead) and a relatively high $1.5$MB communication overhead at each protocol round (with $72$,$704mJ$). 

However, as discussed in Section~\ref{sec:aggr}, we can use Count-Min Sketches to reduce complexities from linear to logarithmic in the size of the input. That is, we can compress the $582 \times 582$ origin-destination matrix into one of size $272 \times 11$, yielding $4.6s$ computation (with $461mJ$), 
and $11.6KB$ (with $823mJ$) overhead when setting $\epsilon$ and $\delta$ parameters to $0.01$. 
Similarly, for the SFC dataset, an origin-destination matrix would become very large, i.e., $10^4 \times 10^4$. Encrypting such a matrix would yield order of $1$,$000s$ computation overhead (and $1$,$000J$) and transferring it would introduce a $39MB$  overhead (and $2$,$835J$). Obviously, these numbers are are prohibitive large for a mobile application. Whereas, with succinct data representation the matrix size could reduce to $1.9MB$ (with $138$,$137mJ$ transmission overhead), and a computation overhead of $18s$ ($5$,$422mJ$) when using a Count-Min Sketch with $\epsilon, \delta$ parameters both set to $0.01$.

\descr{Accuracy.} Finally, we assess the accuracy loss introduced by the succinct data representation. %
We start by measuring the error in the aggregate counts on the TFL data, even though, as discussed above, 
the size of the input is sufficiently small that we do not actually need input compression. Regardless, we do so to
determine the viability of using Count-Min Sketches, i.e., whether or not the error they introduce would reduce the effectiveness of the analytics, and the results are very encouraging. When setting the $\epsilon, \delta$ parameters to the standard $0.01$, we notice that the overall accuracy error in the resulting aggregate counts, over all time slots and all stations, is in the order of $1\%$. We also measure how the error ``propagates'' on the predictive analytics and, again, do not observe any statistically significant difference: for instance, when we make traffic forecasts for the 100 busiest stations on March 25 (as done in Section~\ref{sec:predictions}), we measure the average forecast error as $60.81 \pm 39.62$ using Count-Min Sketches and $59.53 \pm 42.48$ without it.

\section{Related Work}	
\label{sec:related}
\noindent\textbf{Mobility Analytics.} Zhong et al.~\cite{zhong2016variability} propose a metric for capturing variability in commuting trips and analyze urban mobility patterns in London, Singapore, and Beijing, using one-week data of underground journeys. They show that regularity is exhibited when considering time intervals longer than 15 minutes, and demonstrate that peak hours are those with the least variability during a day. Although their analysis results provide useful insights for our work, the authors do not present any methodology for predicting mobility. 
Silva et al.~\cite{silva2015predicting} introduce a general framework for predicting traffic volumes in the London underground: they build a predictive model for each pair of stations under normal conditions (called the natural regime) and then extend it to model disruptions like station or line closures. Their approach is substantially different from ours since disruptions are actually part of a ground truth dataset they obtain from London's transport authority.

Horvitz et al.~\cite{horvitz2012prediction} propose JamBayes, a probabilistic traffic forecasting system deployed in the Seattle Greater Area. They collect %
highway traffic data and contextual data (e.g., city events), and use Bayesian structure search to model bottlenecks. Additionally, using data of historic traffic \textit{surprises}, in combination with recent data before an event, their system learns Bayesian networks that infer the likelihood of a future surprise event. Garz{\'o} et al.~\cite{garzo2013real} use distributed streaming algorithms to process large scale mobility data and make user mobility predictions on large metropolitan areas. They evaluate their location prediction methods on a 2-week mobility dataset obtained from the Orange D4D challenge~\cite{blondel2012data}. 
Yava{\c{s}} et al.~\cite{yavacs2005data} use data mining to predict user movements in a mobile computing system and evaluate their algorithms on a simulated mobidity dataset. %
Overall, prior work on mobility analytics differ from ours as they do not consider collecting data directly from users (nor the privacy implications thereof).

\descr{Detecting Traffic Anomalies.} %
In~\cite{pan2013crowd}, Pan et al.~combine mobility data along with social media to uncover the road network sub-graph associated with an anomaly, based on the routing behavior of drivers. Their system is evaluated on the Beijing taxi traces dataset. 
Similarly, Thom et al.~\cite{thom2012spatiotemporal} present a system geared to detect spatio-temporal anomalies by performing clustering on geolocated Twitter messages and visualize them using tag clouds. They experiment with three case-studies: an earthquake on the US East Coast, London riots, and hurricane Irene. Barria et al.~\cite{barria2011detection} present an anomaly detection algorithm for road traffic using microscopic traffic variables like relative speed of vehicles, inter-vehicle time gap, and lane changing, and evaluate their approach using real-world video images from a highway segment in Bangkok. Zheng et al.~\cite{zheng2015detecting} investigate whether collective detection of anomalies from multiple spatio-temporal datasets is possible. They propose a probabilistic anomaly detection method based on a spatio-temporal likelihood ratio test and evaluate it on five datasets from New York City.
Sun et al.~\cite{sun2004mobility} build Markov models on user mobility patterns in a cellular network, aiming to detect intrusions in the network. 
Note that, although we also focus on identifying event mobility anomalies, unlike these works, we do so using aggregate crowd-sourced location data -- specifically, collected directly from users in a privacy-preserving way. 

\descr{Private Statistics.} Prior work has also proposed a number of tools to privately gather location statistics, however, they do not demonstrate how these statistics can actually be used for performing mobility analytics, which is one of our main goals.
Ho et al.~\cite{ho2011differential} apply differential privacy to discover interesting geographic locations on aggregate location data, whereas~\cite{machanavajjhala2008privacy} relies on synthetic data generation for publishing statistical information about commuting patterns.
Brown et al.~\cite{brown2013haze} propose Haze, a system for privacy-preserving real time traffic statistics based on jury voting protocols and differential privacy. Their system hides individual data while allowing aggregate information to be collected at the service provider. Similarly, Popa et al.~\cite{popa2011privacy} present PrivStats, a system for computing aggregate statistics over location data achieving privacy and accountability. %
Kopp et al.~\cite{kopp2012privacy} also propose a framework enabling the collection of quantitative visits to sets of locations following a distributed approach.
Shi et al.~\cite{shi2011privacy} show how an untrusted data aggregator can learn statistics over multiple participants' private data using cryptographic techniques along with a data randomization procedure for achieving distributed differential privacy, while Melis et al.~\cite{melis} demonstrate how to combine privacy-preserving aggregation with succinct data structures (Count-Min Sketches~\cite{cormode2005improved}) to efficiently compute statistics whilst provably protecting privacy of single data points. They also consider aggregating location information as a possible application of their protocols but do not perform any analytics. PASTE~\cite{rastogi2010differentially} introduces a solution in a similar setting whereby distributed differential privacy is used on time series data using a Fourier perturbation algorithm.

Finally, Fan et al.~\cite{fan2012real} propose FAST, an adaptive system for releasing real-time aggregate statistics with differential privacy. Their approach is based on a \textit{trusted} central authority that adaptively samples the time series according to detected data dynamics to minimize the overall privacy budget. They employ Kalman filtering to predict data at non-sampling points and estimate the true values from perturbed ones at sampling in order to improve the accuracy of data release. 
In follow-up work~\cite{fan2013differentially}, they present a generic differentially-private framework for anomaly detection on  aggregate statistics, focusing on detecting epidemic outbreak: real-time aggregate data is perturbed using FAST~\cite{fan2012real} and released to an untrusted entity that performs the anomaly detection task. Whereas, we do not use  differential privacy to protect users' privacy, as this would require the presence of a trusted aggregator and introduce a trade-off between privacy and utility that is challenging to tune.

\section{Conclusion}
\label{sec:concl}

This work investigated the feasibility of performing crowd-sourced mobility analytics over aggregate location data, in a setting where users periodically report locations to a server, in such a way that the server can only recover aggregates, thanks to the use of a privacy-preserving aggregation protocol. We experimented with real-world mobility datasets obtained from the Transport For London authority as well as the San Francisco Cabs network, and demonstrated that aggregate location data can be useful for predictive analytic tasks like forecasting traffic volumes in \textit{regions of interest} (ROIs) and detecting anomalies in them, using a methodology based on time series modeling with seasonality. In the presence of traffic anomalies, we also showed how to enhance their traffic volume predictions using additional information from correlated ROIs. Finally, we proposed a privacy-respecting system for data collection, and prototyped a mobile application -- Mobility Data Donors (MDD) --  which we empirically evaluated in terms of computation, communication, and energy overhead.

As part of future work, we plan to evaluate our methodology on different location datasets as well as perform a thorough (differential) privacy analysis of releasing datasets composed of aggregate locations, focusing on group sizes and semantic characteristics of ROIs such as size and density, and their evolution over time.

\descr{Acknowledgments.} The authors wish to thank Luca Melis, George Danezis, Mirco Musolesi, and Licia Capra for their useful feedback and/or help with the experiments and the datasets. This research is supported by a Xerox University Affairs Committee grant on ``Secure Collaborative Analytics.''

%
%\bibliographystyle{abbrv}
%\bibliography{bibfile}

\end{document}